\documentclass[prd,twocolumn,reprint,amsmath,amssymb,superscriptaddress,longbibliography,nofootinbib]{revtex4-1}
\usepackage{bm}
\usepackage{dsfont}
\usepackage[usenames,dvipsnames]{xcolor}
\usepackage{pstricks}
\usepackage[tight]{subfigure}
\usepackage{verbatim}
\usepackage{units}
\usepackage{multirow}
\usepackage{enumitem}
\usepackage{mathrsfs}
\usepackage{leftidx}
\usepackage{xspace}
\usepackage{pifont}
\usepackage{bbm}
\usepackage{epsfig}
\usepackage{natbib}
\usepackage{hyperref}
\hypersetup{
	colorlinks=true,
	linktoc=all,
	linkcolor=blue,
	breaklinks=true,
	citecolor=blue,
	urlcolor=blue}
\usepackage[utf8]{inputenc}
\usepackage[english]{babel}
\newcommand{\e}{\varepsilon}
\newcommand{\s}{\sigma}
\newcommand{\up}{\uparrow}
\newcommand{\down}{\downarrow}
\newcommand{\w}{\omega}

\newcommand{\ket}[1]{\left| {#1} \right\rangle }
\newcommand{\GF}{\mathcal{G}}

\renewcommand{\Im}{\mathrm{Im}\xspace}
\newcommand{\SMM}{\mathcal{S}}

\newcommand{\vecg}[1]{\pmb{#1}}

\newcommand{\vopSc}{\skew{3}{\hat}{\bm{S}}_{\text{c}}}

\newcommand{\Sc}{S_{\text{c}}}
\newcommand{\vopSml}{\skew{2}{\hat}{\bm{s}}}

\newcommand{\vopSt}{\skew{3}{\hat}{\bm{\mathcal{S}}}}

\newcommand{\opStz}{\skew{3}{\hat}{\mathcal{S}}_{\!z}}
\newcommand{\St}{\mathcal{S}}
\newcommand{\avStz}{\langle\mathcal{S}_z\rangle}
\newcommand{\opQzz}{\skew{3}{\hat}{\mathcal{Q}}_{zz}}
\newcommand{\avQzz}{\langle \mathcal{Q}_{zz}\rangle}

\newcommand{\Hameff}{\skew{3}{\hat}{\mathcal{H}}_\text{eff}}
\newcommand{\HamML}{\skew{3}{\hat}{\mathcal{H}}_{\text{ML}}}
\newcommand{\Hammol}{\skew{3}{\hat}{\mathcal{H}}_{\text{mol}}}
\newcommand{\HamT}{\skew{3}{\hat}{\mathcal{H}}_{\text{T}}}
\newcommand{\Hamel}{\skew{3}{\hat}{\mathcal{H}}_{\text{leads}}}
\newcommand{\opn}{\skew{1}{\hat}{n}}
\newcommand{\opd}{\skew{3}{\hat}{d}}
\newcommand{\opc}{\skew{2}{\hat}{c}}

\newcommand{\GS}{\Gamma_{\text{S}}}
\newcommand{\GSc}{\Gamma_{\text{S}}^\ast}
\newcommand{\LL}{\text{L}}
\newcommand{\RR}{\text{R}}
\newcommand{\sfA}{\mathcal{A}}
\newcommand{\eb}{\Delta \mathcal{E}}
\newcommand{\Dth}{\widetilde{D}}
\newcommand{\TK}{T_\text{K}}

\newcommand{\fig}[1]{Fig.\,\ref{fig#1}}
\newcommand{\figs}[1]{Figs.\,\ref{fig#1}}

\newcommand{\etal}{\emph{et al.}\xspace}
\newcommand{\via}{\emph{via}\xspace}
\newcommand{\ie}{\emph{i.e.}\xspace}
\newcommand{\cf}{\emph{cf.}\xspace}
\newcommand{\eg}{\emph{e.g.}\xspace}

\newcommand{\circon}{\protect\raisebox{-1.25pt}{\protect\scalebox{1.3}{\ding{192}}}\xspace}
\newcommand{\circtw}{\protect\raisebox{-1.25pt}{\protect\scalebox{1.3}{\ding{193}}}\xspace}
\newcommand{\circth}{\protect\raisebox{-1.25pt}{\protect\scalebox{1.3}{\ding{194}}}\xspace}

\newcommand{\Eq}[1]{Eq.~(\ref{#1})}
\newcommand{\Fig}[1]{Fig.~\ref{fig:#1}}
\newcommand{\Sec}[1]{Sec.~\ref{sec:#1}}

\newcommand{\EEq}[1]{Eq.~(\ref{eq:#1})}

\newcommand{\EFig}[1]{Fig.~\ref{#1}}

\newcommand{\sA}{\text{a}}
\newcommand{\sB}{\text{b}}
\newcommand{\core}{\text{core}}
\newcommand{\ML}{\text{ML}}
\newcommand{\even}{\text{even}}
\newcommand{\odd}{\text{odd}}
\newcommand{\Stz}{\mathcal{S}_z}
\newcommand{\SzMM}{\mathcal{S}_z}
\newcommand{\Scz}{S_{\text{c},z}}
\newcommand{\tD}{\widetilde{D}}
\newcommand{\Ds}{D_\text{spin}}

\newcommand{\beq}{\begin{equation}}
\newcommand{\eeq}{\end{equation}}
\newcommand{\beqa}{\begin{eqnarray}}
\newcommand{\eeqa}{\end{eqnarray}}

\usepackage[normalem]{ulem}
\setcitestyle{numbers,square}

\begin{document}

\title{Giant superconducting proximity effect on spintronic anisotropy}
\author{Krzysztof P. W{\'o}jcik}
\email{kpwojcik@ifmpan.poznan.pl}
\affiliation{Institute of Molecular Physics, Polish Academy of Sciences, 60-179 Pozna{\'n}, Poland}
\affiliation{Physikalisches Institut, Universit\"{a}t Bonn, Nussallee 12, D-53115 Bonn, Germany}
\author{Maciej Misiorny}
\affiliation{Department of Microtechnology and Nanoscience MC2, Chalmers University of Technology, SE-412 96 G\"{o}teborg, Sweden}
\affiliation{Faculty of Physics, Adam Mickiewicz University, Uniwersytetu Poznańskiego 2, 61-614 Pozna{\'n}, Poland}
\author{Ireneusz Weymann}
\affiliation{Faculty of Physics, Adam Mickiewicz University, Uniwersytetu Poznańskiego 2, 61-614 Pozna{\'n}, Poland}

\date{\today} 
\begin{abstract}
We investigate theoretically the interplay of proximity effects due to the presence of a superconductor 
and normal ferromagnetic leads on the formation of the spintronic quadrupolar exchange field in a large-spin molecule.
We show that the spintronic anisotropy can be enhanced by a few orders of magnitude,
and tuned by changing the strength of coupling to the superconductor.
Especially large anisotropy is generated in the vicinity of the charge parity changing transition of the molecule.
We also provide predictions of measurable spectral properties being the hallmarks of these phenomena.
\end{abstract}

\maketitle

%%%%%%%%%%%%%%%%%%%%%%%
\section{Introduction}
\label{sec:intro}
%%%%%%%%%%%%%%%%%%%%%%%

Devices based on large-spin (\mbox{$\SMM\geqslant1$}) nanoscopic systems (\eg, adatoms~\cite{Jacobson_Nat.Commun.6/2015,Donati16}, molecules~\cite{Mannini09,Vincent12,Natterer17} or engineered quantum-dot structures~\cite{Maune2012Jan,Malinowski2018Mar})
hold promise to become crucial components of electronic and spintronic circuits.
The behavior of a spin-degree ground multiplet in such a device is captured by the effective Hamiltonian~\cite{Boca_book},
\begin{equation}
	\Hameff
	=
	B\opStz+D\opQzz
	.
\end{equation} 
Here, the first term represents the effect of a magnetic field~$B$ on the component of the total spin along the $z$-axis, $\opStz$. The second term, on the other hand, describes uniaxial magnetic anisotropy (quantified by a constant~$D$) associated with a diagonal, $z$th component of the spin-quadrupole moment tensor, 
\begin{equation}
	\opQzz\equiv\opStz^2-\St(\St+1)/3.
	\label{Qzz_def}
\end{equation}
Note that all quantities and parameters are set in units of energy, that is, \mbox{$\hbar=|e|=k_\text{B}=1$}.
Impor\-tant\-ly, for \mbox{$D<0$}, this anisotropy leads to formation of the \emph{energy barrier}~$\eb$ for spin reversal between two metastable states~$\ket{\pm\St_z}$, see \fig{1}(a)---the prerequisite for a \emph{magnetic bistability}, indispensable for applications in information-processing technologies~\cite{Lehmann2009Mar,Aromi2012,Khajetoorians2016Apr}.
In an \emph{electronic} circuit, $B$ corresponds then to an externally applied magnetic field, while $D$  stems from the intrinsic spin-orbit interaction in the device~\cite{Gatteschi_book}.
However, in a \emph{spintronic} circuit, where electronic transport is spin-polarized, both $B$ and $D$ can be entirely generated spintronically in the form of effective dipolar~\cite{MartinekTK,MartinekEx,Hauptmann_Nat.Phys.4/2008,Gaass_Phys.Rev.Lett.107/2011} and quadrupolar~\cite{Misiorny2012Dec,Misiorny2013,YSRanisotropy} exchange fields, respectively---here, also referred to as `spintronic fields'.
As a result, both the magnetic anisotropy of the device and its spin state can be on demand controlled in a fully electrical manner.
The success of this approach hinges on the possibility of tuning and inducing significantly large spin-reversal barriers,
which in the case of intrinsic magnetic anisotropy can be relatively small and hardly fine-controllable~\cite{Sessoli17}.

\begin{figure}[b]
	\includegraphics[width=0.99\columnwidth]{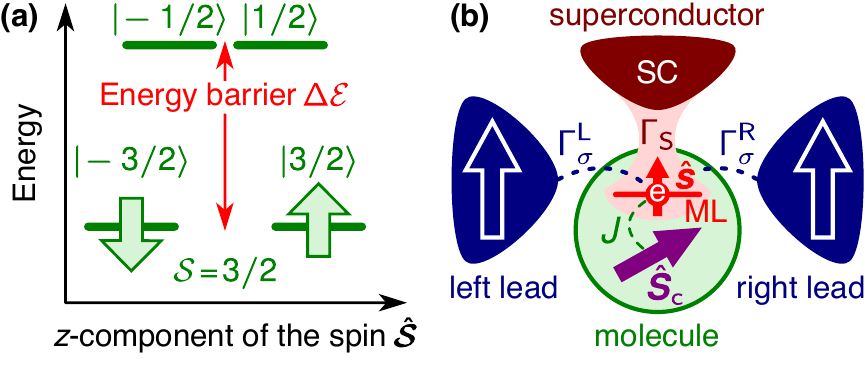}
	\caption{
	(a) Example of an energy spectrum for a model system with the ground spin-multiplet \mbox{$\St=3/2$} and
	uniaxial spintronic anisotropy \mbox{$D<0$}. 	(b) Schematic of a SC-proximized large-spin molecule
	[with a molecular level (ML) occupied by a single electron] connected to two metallic ferromagnetic
	leads. For details see the main text.
	}
	\label{fig1}
\end{figure}

In this paper, we propose how to boost this barrier by orders of magnitude and control it with the aid of the superconductor (SC) proximity effect.
Hybrid nanostructures involving molecules or quantum dots attached
to normal and SC electrodes are nowadays attracting 
a considerable attention \cite{hybridQDs,MartinRodero_AdinPhys2011}. 
Such systems enable the exploration of Yu-Shiba-Rusinov states \cite{Yu,Shiba,Rusinov}
by means of Andreev spectroscopy \cite{Lee2014Jan,vdZant_inducedShiba,YSRscreening}
and have proved useful for solid-state-based generation
of entangled states by splitting the Cooper pairs~\cite{CPS1,CPS9}.
Moreover, the superconducting proximity can strongly influence
the transport properties of hybrid devices, especially in the strong correlation regime,
and generate a quantum critical behavior~\cite{Franke_Science11,Pillet2013Jul,Domanski-IW}.
Here, we focus on a new aspect of SC proximity, namely,
we unveil its beneficial impact on spintronic magnetic anisotropy in molecular systems.
To demonstrate this impact, we consider a prototypical hybrid device consisting of a large-spin, intrinsically isotropic molecule%
\footnote{For the case of a molecule with an intrinsic component of uniaxial magnetic anisotropy see \Sec{D}.}
embedded into a magnetic tunnel junction and proximized by a SC, see \fig{1}(b).
Employing the numerical renormalization group (NRG) approach~\cite{WilsonNRG},
electronic transport through such a device is investigated reliably 
in a broad range of tunnel couplings between the molecule and the junction.
We show that the spintronic anisotropy can be controllably tuned by 
changing the strength of coupling to the superconductor.
In particular, we predict an enhancement of anisotropy by a few orders of magnitude due to the SC proximity
for parameters tuned around the transition between two different ground states of the molecule.
In general, the mechanism underlying such a large enhancement of the induced magnetic anisotropy 
can be attributed to a strong renormalization of Coulomb correlations in the molecule
due to the superconducting proximity effect---for detailed explanation see \Sec{QPT}.

We believe that our work sheds new light on the unexplored aspects of the
interplay of superconducting and ferromagnetic proximity effects
in transport through large spin molecules. In particular,
it opens new perspectives for manipulating the properties of
magnetic molecules, which are indispensable for applications in future information
storage technologies and molecular spintronics.

To make the main message of the paper more evident, we organized it as follows. 
After the introduction, the model and the methods used for its solution
are explained in \Sec{model}. Then, the most relevant consequences of the ferromagnetic and superconducting
proximity effects are discussed in \Sec{Qex} and \Sec{QPT}, respectively, and the results concerning the giant
induced anisotropy are given in \Sec{spFun}. 
Finally, we extend our analysis to a few more general cases and show that the qualitative understanding 
of the results presented in \Sec{spFun} remains valid. Specifically, these cases include different values 
of the molecule's spin (Sec.~\ref{sec:spin}) and the analysis of the example of an intrinsically anisotropic
molecule (Sec.~\ref{sec:D}).
After concluding the paper in \Sec{conclusions} we also extend the main results in the Appendices. 
The exact energies and eigenstates of the system in the limit of a molecule decoupled from the normal 
leads are analyzed in Appendix~\ref{sec:SCmol}.
In Appendix~\ref{sec:sfA} we discuss the local spectral density of the relevant molecular level in the 
full range of energies, commenting the features irrelevant for the analysis of the spintronic anisotropy,
and thus, not addressed in the main text. Finally, in Appendix~\ref{sec:Qzz0} we describe in detail 
the complementary theoretical approach used for calculating the induced anisotropy, 
which does not resort to the examination of the spectral features.

%%%%%%%%%%%%%%%%%%%%%%%
\section{Theoretical description and method}
\label{sec:model}
%%%%%%%%%%%%%%%%%%%%%%%

A large-spin molecule is modeled as composed of~two parts~\cite{Misiorny_vdZant}: a~magnetic core (represented as an effective spin~$\vopSc$) and a single molecular level (ML), which mediates transport of electrons between the leads. 
The ML can be occupied by up to two electrons, which is accounted for by the Hamiltonian
\begin{equation}
	\HamML
	=
	\e\sum_\s\opn_\s+U\opn_\up\opn_\down
	.
\end{equation}
Here, \mbox{$\opn_\s\equiv\opd_\s^\dagger\opd_\s^{}$} is the occupation operator for an electron of spin~$\s$, whereas~$\e$ and~$U$ stand for single-occupation and Coulomb energies, respectively.
Furthermore, it is assumed that if an electron resides in the ML, its spin
\mbox{$\vopSml\equiv (1/2)\sum_{\s\s^\prime}\opd_\sigma^\dagger \vecg{\s}_{\s\s^\prime}\opd_{\s^\prime}^{}$}
[with \mbox{$\vecg{\s}\equiv(\s^x,\s^y,\s^z)$} denoting the Pauli matrices]
couples \via ferromagnetic exchange interaction~\mbox{$J>0$} to the core spin~$\vopSc$, so that the total spin of the molecule reads as~\mbox{$\vopSt\equiv\vopSc+\vopSml$}. 
Finally, the molecule is considered to be proximized by a SC lead [depicted in \fig{1}(b) as a shaded region], with the relevant coupling strength~$\GS$.
Such proximized molecules, impurities or quantum dots have been already extensively studied 
and their properties are qualitatively captured by the effective Hamiltonian in the large-gap limit~\cite{RozhkovArovas,Buitelaar,Domanski-IW,Pawlicki2018Aug}
\begin{equation}
	\label{eq:H_SMM}
	\Hammol=\HamML-J \vopSc\!\cdot\!\vopSml-\GS\big[\opd_{\up}^\dagger\opd_{\down}^\dagger+\text{h.c.}\big]. 
\end{equation}
It describes correctly the emergence of a pairing potential for two electrons occupying ML 
through the last term in~\Eq{eq:H_SMM}, as well as the renormalization of the molecular states and energies resulting from $\GS>0$. It also allows for calculation of Andreev bound states energies; see Appendix~\ref{sec:SCmol},
and even more importantly, reproduces Yu-Shiba-Rusinov impurity physics \cite{YSRscreening}.

Next, the normal leads [\mbox{$q=\LL(\text{eft}),\RR(\text{ight})$}] of the junction
are described as reservoirs of spin-polarized, non-interacting itinerant electrons,
\begin{equation}
	\Hamel
	=
	\sum_{qk\s}\e_{k\s}^{q\phantom{\dagger}}\opc_{k\s}^{q\dagger}\opc_{k\s}^{q\phantom{\dagger}}
	.
\end{equation}
The operator~$\opc_{k\s}^{q\dagger}$ ($\opc_{k\s}^{q}$) creates (annihilates) a spin-$\s$ electron with energy~$\e_{k\s}^{q\phantom{\dagger}}$ and momentum~$k$ in the $q$th lead.
Tunneling of electrons between leads and the molecule is captured by 
\begin{equation}
	\HamT
	=
	\sum_{qk\s}
	\sqrt{\frac{\Gamma_\s^q}{\pi\rho}}
	\,
	\big[\opc_{k\s}^{q\dagger}\opd_{\s}^{}+\text{H.c.}\big]
	,
\end{equation}
where~$\Gamma_\s^q$ denotes the spin-dependent hybridization function. 
In~addition, a symmetric and flat conduction band within the range~\mbox{$[-W,W]$} is assumed in both leads, with the constant density of states~\mbox{$\rho=1/(2W)$}.
\ Assuming the~left-right coupling symmetry,  the tunnel-coupling to two transport leads can be  reduced at equilibrium to a single-channel problem~\cite{even-odd} with a new effective hybridization \mbox{$\Gamma_\s=\Gamma_\s^\LL+\Gamma_\s^\RR$}. 
Since the primary focus of this paper is on the formation of spintronic fields, we consider here only the \emph{parallel} relative orientation of spin moments in the leads.
In such a configuration, one finds \mbox{$\Gamma_{\up(\down)}=\Gamma(1\pm p)$} with \mbox{$\Gamma\equiv\Gamma_\up+\Gamma_\down$}
and $p$ being the spin polarization of the leads.

Spin-resolved transport properties of the device are determined by the normalized spectral function of the ML, \mbox{$\sfA_\s(\w)=- \Gamma_\s\Im\big\{\GF_\s^{\text{r}}(\w)\big\}$},
where $\GF_\s^{\text{r}}(\w)$ is the Fourier transform of the retarded Green's function
$\GF_\s^{\text{r}}(t)=
-i\theta(t)\big\langle\!\big\{\opd_\sigma^{}(t),\opd_\sigma^\dagger(0)\big\}\!\big\rangle$.
This allows us to find the linear-response conductance
between the two ferromagnets, $G=G_\uparrow + G_\downarrow$,
with 
\begin{equation}
	G_\sigma
	=
	\frac{e^2}{h}
	\int d\omega
	\bigg[-\frac{\partial f(\omega)}{\partial\omega}\bigg]
	\sfA_\s(\w)
	,
\end{equation}
where $f(\omega)$ is the Fermi function.
The spectral function together with other averaged relevant quantities are
obtained with the aid of the NRG method~\cite{WilsonNRG,fnrg}.
We use the discretization parameter \mbox{$3\leq \Lambda \leq 4$} and keep at least $N=3000$ states at each iteration.
The conductance and expectation values are calculated directly from discrete NRG data. The spectral functions are obtained using the $z$-averaging trick \cite{Z} with two discretization meshes.

%%%%%%%%%%%%%%%%%%%%%%%
\section{Quadrupolar spintronic field}
\label{sec:Qex}
%%%%%%%%%%%%%%%%%%%%%%%

Generally, \mbox{$\avStz\neq0$} implies that a (large) spin is subject to a magnetic field, either 
real (externally applied) or spintronic one, while \mbox{$\avQzz\neq0$} indicates the presence of
an uniaxial magnetic anisotropy, regardless of its origin.
Since the main focus is here on the quadrupolar field, we assume henceforth that the molecular level
is electrically tuned to the particle-hole symmetry point~(\mbox{$\e=-U/2$}),
where the dipolar field disappears and \mbox{$\avStz=0$} \cite{MartinekTK}.

\begin{figure}[bt]
\includegraphics[width=0.99\columnwidth]{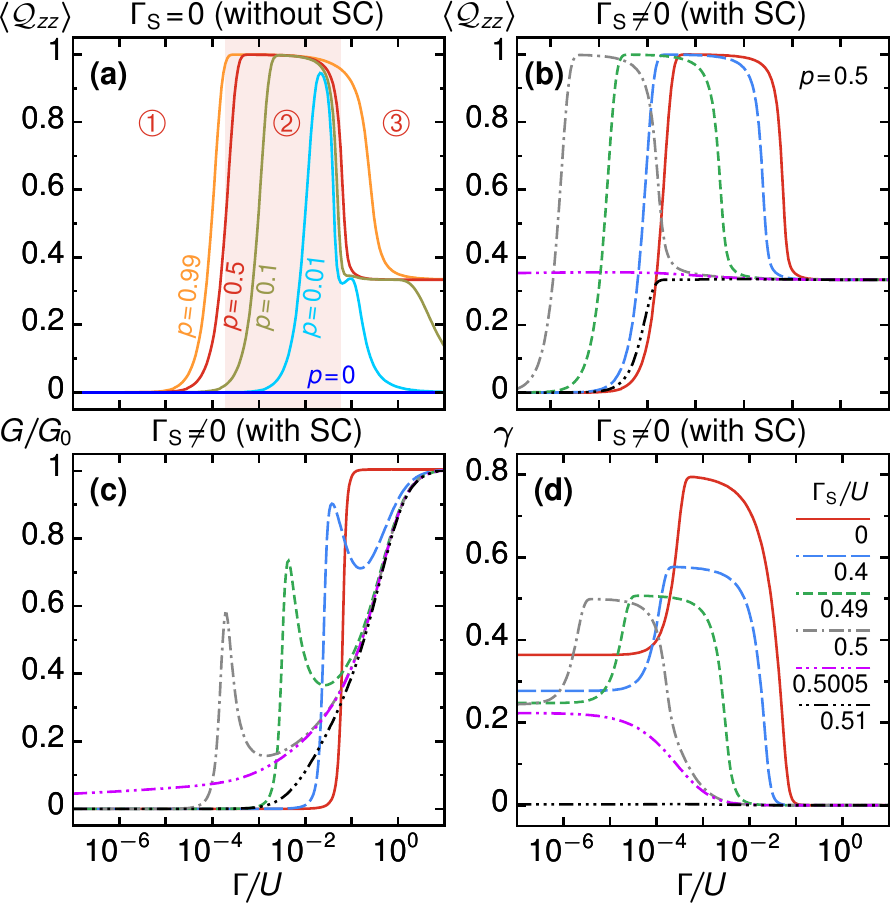}
\caption{ 
	(a)~Spin-quadrupole moment~$\avQzz$ for a molecule of spin \mbox{$\SMM=3/2$}  plotted as a function of tunnel coupling~$\Gamma$ for \mbox{$\GS=0$} and several spin polarizations~$p$ of the leads.
	(b)~$\avQzz$ shown for different values of~$\GS$ indicated in (d) and \mbox{$p=0.5$}.
	Panels~(c) and~(d) present the linear-response conductance~$G$ (normalized to \mbox{$G_0\equiv2e^2/h$}) and the spin polarization of trans\-ported electrons~$\gamma$, respectively, which correspond to curves shown in~(b).  
	Other parameters: \mbox{$U/W=0.5$}, \mbox{$J/U=10^{-3}$} and \mbox{$T/U=10^{-10}$}. 
		 }
\label{fig2}
\end{figure}

In \fig{2}(a) we show for a model molecule with \mbox{$\SMM=3/2$} (\ie, \mbox{$\Sc=1$})
how $\avQzz$ evolves when shifting between the regimes of weak (\circon) and strong (\circth) tunnel coupling~$\Gamma$. For \mbox{$p=0$}, the $SU(2)$ spin symmetry guarantees \mbox{$\avQzz =0$}. However, as soon as \mbox{$p\neq0$}, the quadrupolar spintronic field (\mbox{$D<0$}) becomes active and the ground-state doublet \mbox{$\ket{\pm\St_z}$} gets stabilized due to the onset of the energy barrier~\mbox{$\eb=(2\St-1)|D|$} for spin reversal, see \fig{1}(a). 
It can be seen that  $\avQzz$ saturates to two different values for \mbox{$|D|\gg T$}: 
\mbox{$\avQzz\rightarrow\St(2\St-1)/3=1$}
in the intermediate tunnel-coupling regime~(\circtw), and 
\mbox{$\avQzz\rightarrow\Sc(2\Sc-1)/3=1/3$}
in the strong coupling regime~(\circth).
The reduction of $\avQzz$ in the latter case stems from the arrival of the Kondo correlations,
which manifest through unitary conductance through the device, see the solid line in \fig{2}(c). 
In the Kondo effect the molecular level spin~$\vopSml$
becomes screened by the conduction electrons of metallic leads
\cite{Hewson_book}. This effectively results in the reduction of the total spin of the molecule
to $\St \to \Sc$, which yields \mbox{$\avQzz=1/3$}.
Note that for \mbox{$\St=1$} considered in Ref.~\cite{Misiorny2013},
\mbox{$\Sc=1/2$} so that the molecule is never anisotropic in the Kondo regime.
Conversely, for \mbox{$\St \geq 3/2$} the magnetic core of the molecule remains anisotropic in the Kondo regime (\circth), leading to finite $\avQzz$, clearly visible in \fig{2}(a). However, due to the screening of the molecular level spin,
it is difficult to address and measure the spin state of the molecule's core
by means of only transport spectroscopy of the ML.

The three regimes of tunnel-coupling visible in \fig{2}(a) persist for a wide range of spin polarizations~$p$ of the leads.
Indeed, already for $p=0.01$ a narrow range of \mbox{$\avQzz=1$}
for intermediate couplings \circtw can be observed---meaning that
a high spin polarization of a ferromagnet is not a necessity for generating spintronic anisotropy.
For larger spin polarizations this region broadens quickly
at the expense of the weak coupling region \circon,
due to a quadratic dependence of the induced anisotropy~$D$ on~$p$, $D\propto p^2\Gamma^2$~\cite{Misiorny2013}.
It also expands against the regime of strong coupling~\circth
 as the Kondo temperature decreases with raising the spin polarization of the ferromagnets~\cite{MartinekTK}.
 The latter mechanism is, however, efficient mainly for very large spin polarization,
 see the line for \mbox{$p=0.99$} in \fig{2}(a).

Another hallmark feature of  the quadrupolar field is the enhanced spin-filtering of electrons 
tunneling through the device, quantified by 
the conductance spin polarization,
\begin{equation}
	\gamma=\frac{G_\up-G_\down}{G_\up+G_\down}
	,
\end{equation}
see solid line in \fig{2}(d).
Finite spin polarization is related to the suppression of
electron-tunneling-induced spin-exchange processes involving the axial spin states~$\ket{\pm\St_z}$.
Analyzing how $\gamma$ depends on $\Gamma$,
one can identify crossovers between all three coupling regimes,
with \mbox{$\gamma\to 0$} indicating the onset of the Kondo effect
where the conductance reaches unitary limit in both spin channels.

%%%%%%%%%%%%%%%%%%%%%%%
\section{The significance of the superconductor}
\label{sec:QPT}
%%%%%%%%%%%%%%%%%%%%%%%

The situation changes substantially if the molecule becomes proximized
by superconductor (\mbox{$\GS\neq0$}), \cf solid and dashed curves in \figs{2}(b)-(d). 
First of all, one observes that for finite coupling to superconductor
the quadrupolar field sets in already at a much weaker tunnel coupling to ferromagnetic leads,
\ie, for values of $\Gamma$ few orders of magnitude smaller as compared to the case of \mbox{$\GS=0$}, see \fig{2}(b).
Then, at some critical value of $\GS$, \mbox{$\GSc/U = 0.5005$},
the three different coupling regimes cease to be visible in $\avQzz(\Gamma)$,
and \mbox{$\avQzz \approx 1/3$} in the full range of~$\Gamma$. 
For yet larger $\GS$,~\mbox{$\GS>\GSc$}, only two distinctive regimes emerge:
The first one (for small values of~$\Gamma$), when the molecule is practically
spin-isotropic with \mbox{$\avQzz=0$}, and the second regime (for larger values of~$\Gamma$),
when magnetic anisotropy appears, but $\avQzz$ does not exceed $1/3$, see \fig{2}(b).

In order to understand this behavior, let us first discuss qualitatively what are the consequences of coupling of the molecule to the superconducting lead (even in the absence of the ferromagnetic one).
As already explained in \Sec{model}, the system under consideration can be described by the proximized molecule Hamiltonian given in \Eq{eq:H_SMM}. It can be diagonalized exactly, which is done in Appendix~\ref{sec:SCmol}. 
By comparing \Eq{energies_oo} with \Eq{energies_ee} one can clearly conclude that the proximization of the molecule by a superconductor leads to 
the splitting of excited states energies and consequently to a
decrease of the excitation energy to 
the relevant
states of different charge parity---note that for~\mbox{$\GS\neq0$} the charge itself is not a good quantum number.
Importantly, such a decrease of the excitation energies can also be understood intuitively as an effective reduction of the Coulomb repulsion by the superconducting correlations, which occurs due to mixing of empty and doubly occupied states of the molecular level.
As a result, by reducing the excitation energy, the coupling to superconductor tends to enhance the charge fluctuations of the molecule.

On the other hand, it is important to recall that the spintronic anisotropy is generated by ferromagnetic-proximity effect due to spin-resolved charge fluctuations between the molecule and ferromagnets. 
Specifically, the quadrupolar field can be understood perturbatively (in the second order in $p\Gamma$; for detailed discussion see \cite{Misiorny2013}) as the consequence of correlated charge fluctuations. 
For this reason, the enhancement of the letter due to the superconductor proximity implies that one should also expect the boost in the magnitude of the quadrupolar field.
What makes this boost huge is, however, the existence of the (approximate) critical point, at which the relevant excitation energy vanishes, and thus, the effect of enhancing the spintronic anisotropy is driven to its extreme.

To be precise, the SC-proximized molecule
isolated from normal leads experiences a quantum phase transition driven by increase of $\GS$
at \mbox{$\GSc=(U+JS_{\rm c})/2$},%
\footnote{For parameters used in~\fig{2} one finds  \mbox{$\GSc/U=0.5005$}; see also Appendix~\ref{sec:SCmol} for further explanation.}
which is analogous to that predicted for single \cite{Bauer2007Nov,Domanski-IW}
or double quantum dots \cite{Sherman2016Nov,KWIW-S2QD}.
Then, the ground spin-multiplet changes from \mbox{$\St=3/2$} with oddly occupied level of the molecule,
for \mbox{$\GS<\GSc$}, to two  spin-triplets (\mbox{$\St=1$}), with the even occupation of ML,
for \mbox{$\GS > \GSc$}---implying the reduction of the molecule's spin to \mbox{$\St=\Sc$}.
This phenomenon, known as \emph{Yu-Shiba-Rusinov} (YSR) \emph{screening}~\cite{YSRscreening},
explains the absence of the regime with \mbox{$\avQzz=1$} in \fig{2}(b)
and the complete suppression of the spin-filtering effect visible in \fig{2}(d) for \mbox{$\GS /U=0.51$},
see also Appendix~\ref{app:ABS}.
Importantly, note that this is \emph{not} the Kondo screening, so that the strong tunnel-coupling regime 
for \mbox{$\GS>\GSc$} does \emph{not} correspond to the Kondo regime anymore. 

In the vicinity of the critical point, even very small values of the coupling~$\Gamma$
to the normal leads allow for excitations between the two spin-multiplets.
For \mbox{$\GS=\GSc$}, this results in induced anisotropy in the full tunnel-coupling range [see double-dotted line in~\fig{2}(b)], and a  residual spin-polarized conductance in the weak tunnel-coupling regime [\figs{2}(c)-\ref{fig2}(d)]. 
As long as \mbox{$\GS<\GSc$}, a resonance appears in the conductance whenever $\Gamma$ becomes of the order of \mbox{$\GSc-\GS$}, see \fig{2}(c). More precisely, this resonance starts forming when the broadening
of molecule's energy levels becomes smaller than
the excitation energy between the even and odd ML occupation states, \mbox{$\Gamma\lesssim\GSc-\GS$}.
Then, the system enters the~Kondo regime, which is visible as an enhanced conductance through the system.
However, the conductance quickly drops again with lowering~$\Gamma$
due to exponential dependence of the Kondo temperature~$\TK$
on~$\Gamma$ and the fact that the actual temperature $T$ used in \fig{2},
albeit finite, is very small. Thus, once \mbox{$\TK<T$},
the conductance becomes suppressed with decreasing $\Gamma$,
which is a direct manifestation of the Coulomb blockade effect.
Clearly, the Kondo peak does not arise for \mbox{$\GS\geq \GSc$},
since after crossing the parity-changing transition the molecule is in the YSR-screened ground state. 

\begin{figure}[t]
\includegraphics[width=0.99\columnwidth]{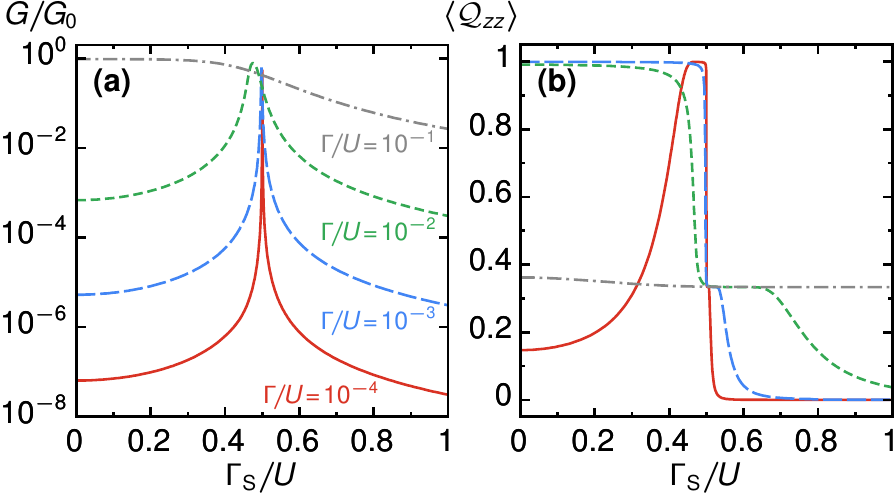}
\caption{ 
	Dependence of (a)~normalized conductance~$G/G_0$ and (b) spin-quadrupole moment~$\avQzz$
	on the coupling~$\GS$, representing the SC-proximity effect on the molecule,
	for selected values of the tunnel coupling~$\Gamma$. 
	Other parameters as in \fig{2}. 
		 }
\label{fig3}
\end{figure}

This phase transition is sharp only in the limit of vanishing coupling
to normal contacts \mbox{$\Gamma\to 0$},
and becomes a smeared crossover for finite $\Gamma$ \cite{Domanski-IW}---observe the increasing width of the corresponding resonance in $G(\GS)$ at \mbox{$\GS\approx\GSc$} in \fig{3}(a). 
Moreover, as can be seen in \fig{3}(a), $\GSc$ gets diminished for larger $\Gamma$,
which explains the suppression of conductance spin polarization $\gamma$ for $\GS$
equal to \mbox{$\GSc=0.5005U$} in the strong coupling 
regime, see the double-dotted line in \fig{2}(d).

Finally, providing complementary information to $G$,
also $\avQzz$ is the convenient quantity for describing the critical behavior of the system, see \fig{3}(b).
The quadrupolar moment changes abruptly at the transition point for
small values of the coupling $\Gamma$,
dropping from \mbox{$\avQzz=1$} (characteristic of \mbox{$\St=3/2$} and \mbox{$D<0$})
to \mbox{$\avQzz=1/3$} (typical for \mbox{$\St=1$}) at \mbox{$\GS=\GSc$}.
On the other hand, for \mbox{$\GS>\GSc$},
$\avQzz$ can decrease yet further owing to the suppression of
the magnetic core anisotropy in the YSR-screened phase. 
Still, relatively strong coupling to ferromagnetic contacts
allows the molecular core to remain anisotropic even
deep in the YSR phase, similar to the Kondo phase for \mbox{$\GS < \GSc$},
see the dot-dashed curve in \fig{3}(b).

%%%%%%%%%%%%%%%%%%%%%%%
\section{Spectral signatures of the SC-proximity}
\label{sec:spFun}
%%%%%%%%%%%%%%%%%%%%%%%

\begin{figure}[t]
\includegraphics[width=0.99\columnwidth]{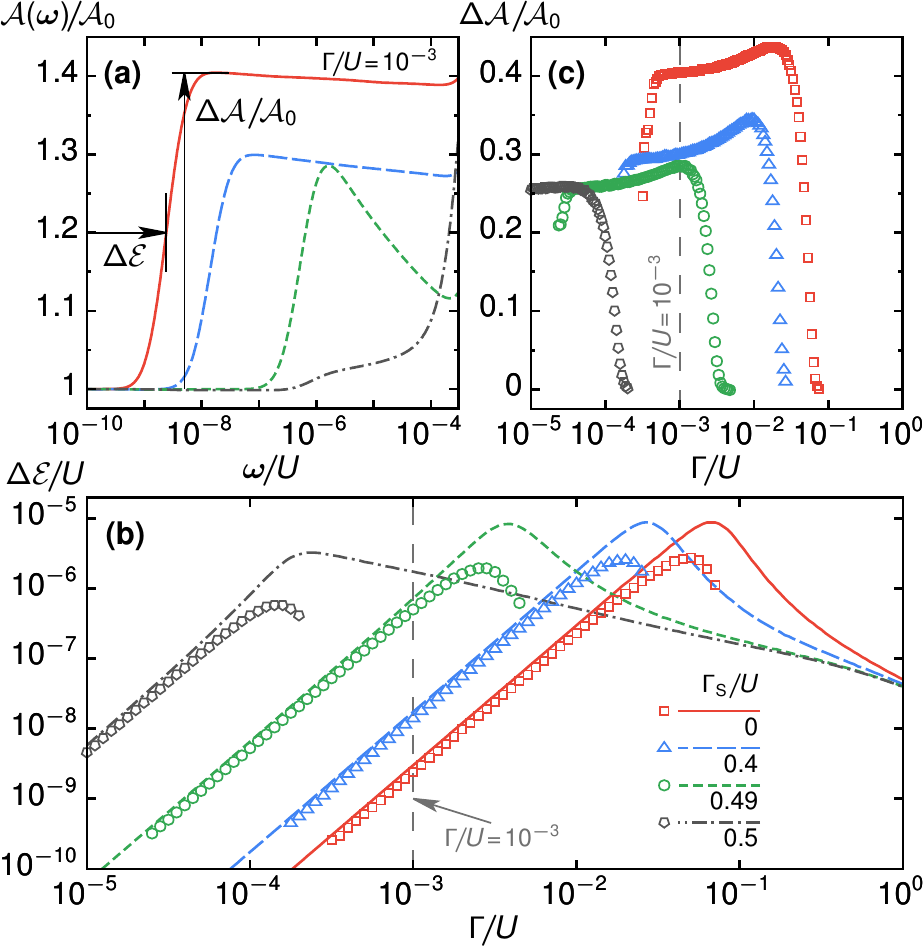}
\caption{ 
	\hspace*{-5pt}
	(a)~Spectral function $\sfA(\omega)$ [scaled by \mbox{$\sfA_0\!\equiv\!\sfA(\omega\!=\!0)$}]
	shown as a function of energy~$\omega$ for selected values of~$\GS$ [given in~(b)] and \mbox{$\Gamma/U=10^{-3}$}.
	All spectral functions are symmetric with respect to \mbox{$\omega=0$}. 
	(b)~Evolution of the excitation gap~$\eb$ with~$\Gamma$ (note a log-log scale).
	\emph{Points} represent~$\eb$ estimated from the inflection point of the step in~$\sfA(\omega)$
	[as, for example, indicated in~(a)], whereas \emph{lines} correspond to~$\eb$ obtained
	using a complementary theoretical method based on the analysis of~$\avQzz$; see Appendix~\ref{sec:Qzz0}.
	The resolution limit for~$\eb$ is set here by thermal energy~$T$. 
	(c)~The height~$\Delta\sfA/\sfA_0$ of the step in~$\sfA(\omega)/\sfA_0$
	used for estimating~$\eb$ in (b) plotted as a function of~$\Gamma$.
	All other parameters as in \fig{2}.
	 }
\label{fig4}
\end{figure}

In experiments, the quadrupolar spintronic field can be accessed indirectly
\via transport measurements of the spectral properties of a device~\cite{Ternes15}.
Particularly, as mentioned above, this field leads to the occurrence of an excitation
gap~$\eb$ between the ground-state doublet~$\ket{\pm\St_z}$ and
the first excited doublet~$\ket{\pm\St_z\mp1}$, see \fig{1}(a).
This excitation gap manifests itself as a pronounced step in the spectral function $\sfA(\omega)$, see \fig{4}(a). 
It is evident that the gap becomes larger by orders of magnitude as~$\GS$
increases from \mbox{$\GS=0$} to its critical value~$\GSc$.
Then, in the vicinity of the critical point, the step in~$\sfA(\omega)$
becomes indistinguishable because the molecule is almost fully conducting, \ie, \mbox{$\sfA_0\equiv\sfA(\omega=0)\approx2$},
with 2 signifying two spin transport channels.
Moreover, for \mbox{$\GS>\GSc$}, the YSR screening becomes
important and magnetic anisotropy is suppressed.
% the above statement is true on in the strong coupling regime !

To investigate the dependence of the induced gap~$\eb$
on the strength of pairing correlations $\GS$ in a wide range of~$\Gamma$,
we estimate~$\eb$ from the inflection point of the step in~$\sfA(\omega)$,
as schematically illustrated in \fig{4}(a). The result is plotted as points in \fig{4}(b),
with the relative hight~$\Delta\sfA$  of corresponding steps used in calculations [see \fig{4}(a)] shown in \fig{4}(c).
One observes that~$\eb$---and thus, the spintronic anisotropy---can be enhanced
by a few orders of magnitude by proximizing the molecule with a superconductor.
To corroborate this finding, in \fig{4}(b) we also present results
for~$\eb$ (lines) derived with another method that does not
resort to the analysis of spectral features. Specifically, in this theoretical approach,
one seeks for some hypothetical component of magnetic anisotropy \mbox{$\Dth>0$}
one would need to add to compensate the quadrupolar spintronic field \mbox{$D<0$}.
It corresponds to numerically (by means of NRG) finding $\Dth$
from the condition \mbox{$\avQzz(\Dth)=0$}, and \mbox{$\eb=(2\St-1)\Dth$};
see Appendix \ref{sec:Qzz0} for details of this method. 
The results obtained with these two methods agree.
In fact, with the latter approach one can address the large-$\Gamma$
regime where electronic transport is dominated by Kondo fluctuations,
so that the examination of the step in $\sfA(\omega)$ becomes impossible, see \fig{4}(c).

The mechanism leading to the enhancement of spintronic anisotropy by proximity effect 
is associated with a strong renormalization of Coulomb correlations in the molecule,
as explained in Sec.~\ref{sec:QPT}.
Strong coupling to superconductor mixes the empty and doubly occupied states,
increasing the charge fluctuations responsible for the development of $\eb$.
These fluctuations grow with increasing $\Gamma$, resulting in $\Gamma^2$-dependence
of the induced anisotropy, which is clearly visible in the weak and intermediate coupling regimes.
However, when with augmenting $\Gamma$ the Kondo effect sets in, 
the \mbox{$\eb \sim p^2 \Gamma^2$} dependence is no longer valid
and $\eb$ starts decreasing with further increase of $\Gamma$.
Finally, the asymptotic behavior of $\eb(\Gamma)$ for \mbox{$\Gamma/U\rightarrow1$}
is the same regardless of~$\GS$, see \fig{4}(b), since tunnel broadening
of all molecular states is so large that all states become strongly hybridized.

%======================================================================
\section{Different values of magnetic molecule spin}
\label{sec:spin}
%======================================================================

%++++++++++++++++++++++++++++++++++++++++++++++++++++++++++++++++++++++
\begin{figure}[tbt]
\includegraphics[width=0.98\columnwidth]{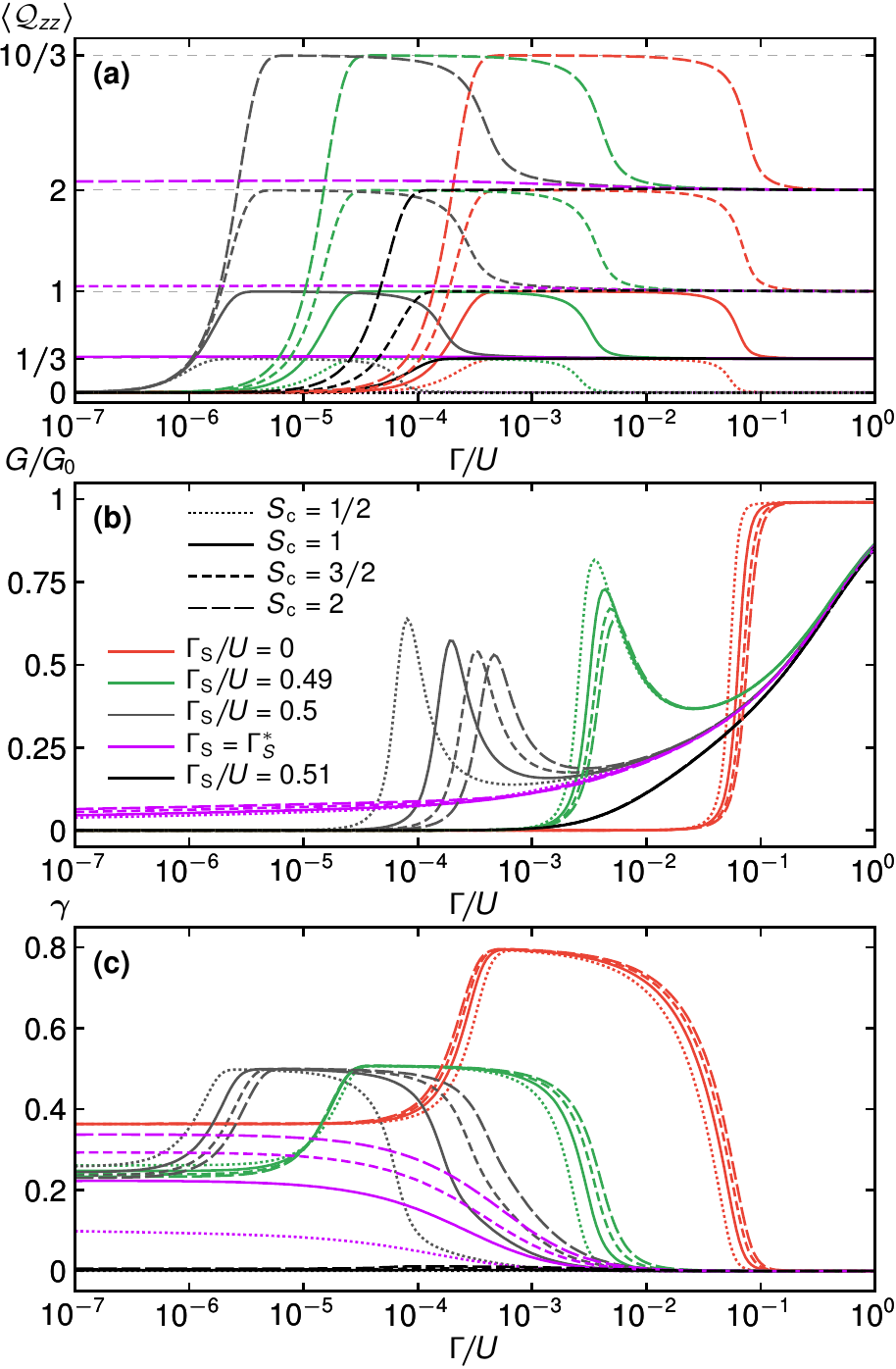}
\caption{
	Extension of \EFig{fig2}(b)-(d) to different values of the magnetic-core spin~$\Sc$. 
	(a)~Spin-quadrupole moment~$\avQzz$ plotted as a function of $\Gamma$ for different values of~$\GS$ (different colors) and $\Sc$ (different styles of dashing).
	Panels~(b) and~(c) present the linear-response conductance~$G$ (as previously normalized to \mbox{$G_0\equiv2e^2/h$}) and the spin polarization of transported electrons~$\gamma$, respectively, which correspond to curves shown in~(a).  
	Other parameters as in \EFig{fig2}.
		 }
\label{fig:S_Qzz}
\end{figure}
%++++++++++++++++++++++++++++++++++++++++++++++++++++++++++++++++++++++

Up to this point we considered a hypothetical molecule characterized by the total spin of \mbox{$\St=3/2$}. In the case of the ferromagnetic exchange coupling~$J$ within the molecule, see \Eq{eq:H_SMM},
and the ML occupied by a single electron, such magnitude of the total spin
is observed for the spin of the magnetic core of the molecule equal to $\Sc=1$.
In this section we discuss the influence of the value of~$\Sc$ on the presented results
and, in particular, on the induced magnetic anisotropy.

First of all, the effect of magnetic anisotropy reveals itself only if \mbox{$\St\geqslant1$},
and thus, we consider only \mbox{$\Sc\geqslant 1/2$}. 
Furthermore, \mbox{$\St=S_c+1/2$} is not valid in all the coupling regimes. This has been discussed 
in the context of \EFig{fig2}(b) and is clearly visible also in \Fig{S_Qzz}(a), where the expectation value 
$\avQzz$ as a function of the tunnel-coupling~$\Gamma$
is presented for different values of the magnetic core spin~$\Sc$ and
the strength of pairing correlations described by~$\GS$.
Let us first discuss the set of curves for~\mbox{$\GS=0$}. Evidently, one observes \mbox{$\avQzz \to 0$} in the limit where the tunnel-coupling is small and it increases to its maximal value,
\mbox{$
	\avQzz 
	= 
	\St_z^2 
	- 
	\St (\St+1)/3
$},
in the intermediate tunnel-coupling regime, where \mbox{$\St_z=\pm \St$} and \mbox{$\St=\Sc+1/2$}. 
By further augmenting~$\Gamma$,  the onset of the Kondo screening in the strong tunnel-coupling regime becomes eventually unavoidable, and it leads to reduction of the effective spin of the molecule to \mbox{$\St = \Sc$}.
Moreover, as discussed in \Sec{QPT}, the increase of $\GS$ drives a transition to the evenly-occupied ML state at \mbox{$\GS=\GSc\equiv (U+\Sc J)/2$}, which also results in effective reduction of the molecule's spin to \mbox{$\St = \Sc$}.
For this reason, it becomes clear that \mbox{$\Sc=1$} is the smallest possible value of~$\Sc$
which guarantees that the effects associated with finite magnetic anisotropy
should be expected to arise in all the considered regimes
of the coupling strength to ferromagnetic leads~$\Gamma$.

In general, as one can see in \Fig{S_Qzz}, the behavior of the system with increasing $\GS$ toward the critical value~$\GSc$ for different values of~$\Sc$ is quantitatively the same as for the case of \mbox{$\Sc=1$} studied so far. 
Still, for a given value of~$\GS$ close to $\GSc$, one can observe some large quantitative differences, visible especially in the position of the resonance in the conductance that is associated with the excitation energy between the oddly and evenly occupied-ML states, see in particular  curves for \mbox{$\GS/U=0.5$} in \Fig{S_Qzz}(b). The shift of this position stems from the dependence of the critical point~$\GSc$ on~$\Sc$
(see Appendix~\ref{sec:SCmol} for details).
Finally, worthy of note is that the spin polarization of transported electrons, $\gamma$, does not  depend significantly on $\Sc$ in the limit of $\Gamma\to 0$, except at the critical point \mbox{$\GS=\GSc$}. This is illustrated in \Fig{S_Qzz}(c).

%++++++++++++++++++++++++++++++++++++++++++++++++++++++++++++++++++++++
\begin{figure}[tb]
\includegraphics[width=0.99\columnwidth]{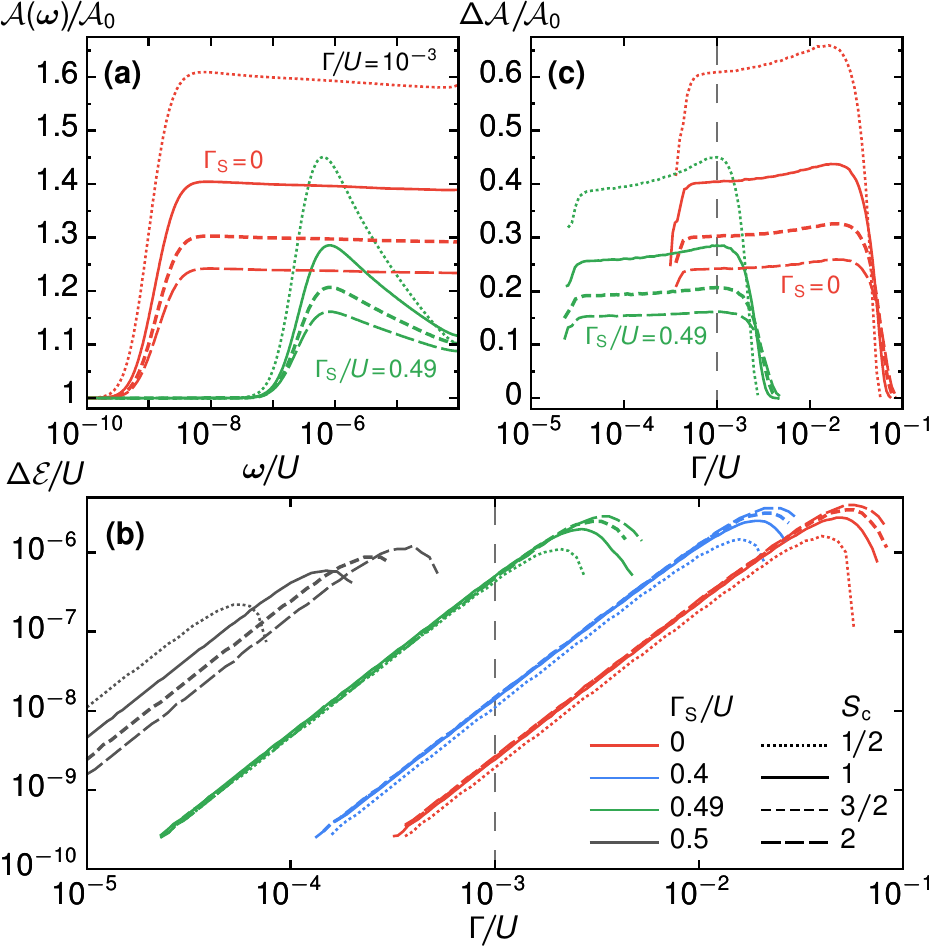}
\caption{
	Extension of \EFig{fig4} to different values of the magnetic-core spin  of the molecule, $\Sc$.
	(a)~Spectral function $\sfA(\w)$ [scaled by \mbox{$\sfA_0\!\equiv\!\sfA(\w\!=\!0)$}] shown as a function of energy~$\w$ for selected values of~$\GS$ and \mbox{$\Gamma/U=10^{-3}$}.
	All spectral functions are symmetric with respect to \mbox{$\w=0$}.
	(b)~Evolution of the excitation gap~$\eb$ with~$\Gamma$ (note a log-log scale). 
	(c)~The height~$\Delta\sfA/\sfA_0$ of the step in~$\sfA(\w)/\sfA_0$ used for estimating~$\eb$ 		in (b) plotted as a function of~$\Gamma$. 
	All other parameters as in \EFig{fig4}.
	 }
\label{fig:S_sfA}
\end{figure}
%++++++++++++++++++++++++++++++++++++++++++++++++++++++++++++++++++++++

The most important point concerning the spin of the molecule is the lack of increase of the energy barrier
with increasing $\St$.
Even though one could naively expect the relation \mbox{$\eb=(2\St -1)|D|$} to hold at least for relatively small $\St$,
in Figs.~\ref{fig:S_sfA}(a)-(b) it is clearly visible that such a dependence is rather very weak.
On the contrary, the energy barrier seems to be hardly affected by changing~$\St$,
\cf Fig.~S-9 in Supplementary Information of Ref.~\cite{Misiorny2013}.
Moreover, in the vicinity of the critical point, where the superconductor
proximity boosts the induced anisotropy most effectively,
the energy barrier~$\eb$ appears to be decreasing for larger values of~$\St$,
see the gray curves in \Fig{S_sfA}(b). This effect is caused by the fact
that the distance from the critical point, \mbox{$\delta\GS \equiv \GSc-\GS$},
is decisive for obtaining the highest anisotropy. Since $\GSc$ increases
with $\Sc$, the largest possible value of the molecular spin~$\St$ is
not necessarily desired for applications, unless arbitrarily
large coupling strengths~$\GS$ are within reach.
A similar conclusion can be drawn from the analysis of \Fig{S_sfA}(c), presenting the height of the step in the spectral function~$\sfA(\w)$ for different values of~$\Sc$. Clearly, the step is the highest for $\Sc=1/2$ and diminishes with increasing $\Sc$, which means that large $\Sc$ is not necessarily the right choice for experiment or applications.

%======================================================================
\section{Results for finite intrinsic anisotropy}
\label{sec:D}
%======================================================================

In this last section we address the effects occurring due to the presence of finite intrinsic magnetic anisotropy (\mbox{$D_0\neq0$}), $\Hammol \mapsto \Hammol + D_0\opStz^2$; see Appendix~\ref{sec:SCmol}
for details. 
Specifically, our goal is to analyze here additional effects that originate from the interplay between the intrinsic ($D_0$) and spintronic ($\Ds$) components of magnetic anisotropy. We note that the key results presented in the previous sections remain qualitatively valid also for finite $D_0$.
We begin the discussion with the analysis of the case when there is no superconductor proximizing the molecule (\mbox{$\GS=0$}), and in the next step we also take into consideration the superconductor-proximity effects.

%----------------------------------------------------------------------
\subsection{Influence of intrinsic magnetic anisotropy  (\mbox{$D_0\neq0$}) for $\GS=0$}
\label{sec:D_GS0}
%----------------------------------------------------------------------

%++++++++++++++++++++++++++++++++++++++++++++++++++++++++++++++++++++++
\begin{figure}[tb]
\includegraphics[width=0.99\columnwidth]{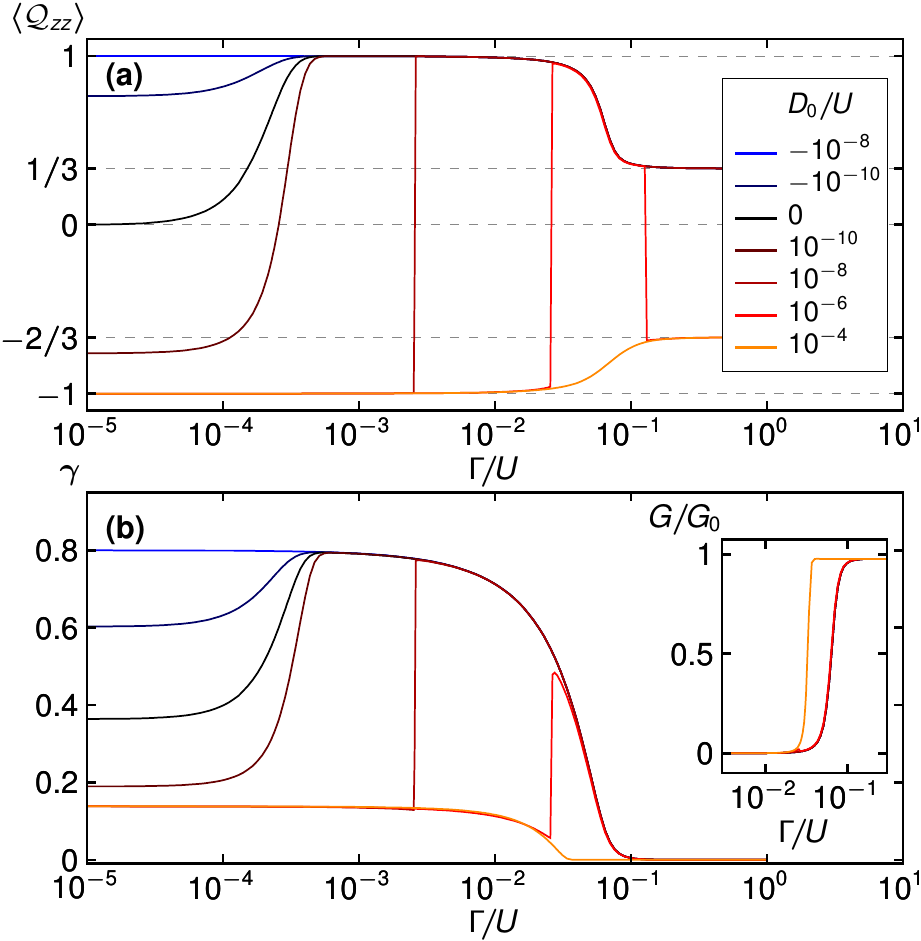}
\caption{
	Generalization of \EFig{fig2}(c)-(d) for \mbox{$\GS=0$} and \mbox{$p=0.5$} to different values of intrinsic magnetic anisotropy of the molecule, $D_0$.
	(a)~Spin-quadrupole moment~$\avQzz$ plotted as a function of $\Gamma$ for different values of~$D_0$.
	Panel~(b) presents the spin polarization of transported electrons~$\gamma$, corresponding to curves shown in~(a).
	Inset in~(b) shows the linear-response conductance~$G$ (normalized to \mbox{$G_0\equiv2e^2/h$}) as a function of $\Gamma$. Other parameters as in \EFig{fig2}.
		 }
\label{fig:Qzz-D}
\end{figure}
%++++++++++++++++++++++++++++++++++++++++++++++++++++++++++++++++++++++

The occurrence of magnetic anisotropy in a large-spin molecule manifests clearly in the dependence of the expectation
value of the $z$-component of the spin-quadrupole moment tensor~$\avQzz$, \Eq{Qzz_def}, on the tunnel-coupling strength~$\Gamma$, as presented in \Fig{Qzz-D}(a). 
One can see in this figure that, unlike for \mbox{$D_0=0$}, in the weak tunnel-coupling limit \mbox{$\avQzz > 0$} for \mbox{$D_0<0$}, while \mbox{$\avQzz < 0$} for \mbox{$D_0>0$}.
In the case of intrinsic  magnetic anisotropy of the `easy-axis' type (\mbox{$D_0<0$})---\ie, when the ground-state spin-doublet \mbox{$\ket{\pm(\Sc+1/2),-}$} is energetically preferable, see Appendix~\ref{sec:SCmol}---one expects that the anisotropic regime should persists also for small tunnel-couplings if only \mbox{$\eb\ll T$} with no other qualitative consequences, which can be observed in \Fig{Qzz-D}.
On the contrary, for \mbox{$D_0>0$} the competition between the intrinsic and spintronic components of magnetic anisotropy becomes apparent. This competition arises as a consequence of the fact that \mbox{$\Ds<0$}, which basically means that the sign of the total effective magnetic anisotropy constant \mbox{$D=D_0+\Ds$} can be tunned by changing~$\Gamma$.
We note that for a half-integer total spin of a molecule~$\St$ (as considered here), positive~$D$ implies that the two-fold degenerate ground state of a minimal $z$-component of spin, \mbox{$\ket{\pm1/2,-}$}, is favored. Such a doublet is in fact \emph{spin-isotropic}, that is, there is no energy barrier (in the absence of magnetic field) for switching between the states of the doublet. 
As a result, the shrinkage or even complete suppression [see the curve for \mbox{$D_0/U=10^{-4}$} in \Fig{Qzz-D}(a)] of the anisotropic regime becomes visible whenever \mbox{$D_0>|\Ds|$}. Moreover, if \mbox{$D_0\gg|\Ds|$}, the Kondo spin-exchange processes within the ground-state doublet get enhanced, which leads to an increase of the Kondo temperature~$\TK$. This behavior can be observed
as a shift of the cross-over value of the coupling strength~$\Gamma$
where the strong-coupling (Kondo) regime sets in towards smaller values,
which is clearly visible in the dependence of the linear response conductance
on~$\Gamma$ for \mbox{$D_0/U=10^{-4}$} presented in the inset of \Fig{Qzz-D}(b).
Interestingly, these effects may be used to tailor the range of the anisotropic regime to a desired width, while the proximity effect of the superconductor can shift it to the desired position.

%++++++++++++++++++++++++++++++++++++++++++++++++++++++++++++++++++++++
\begin{figure}[tb]
\includegraphics[width=0.99\columnwidth]{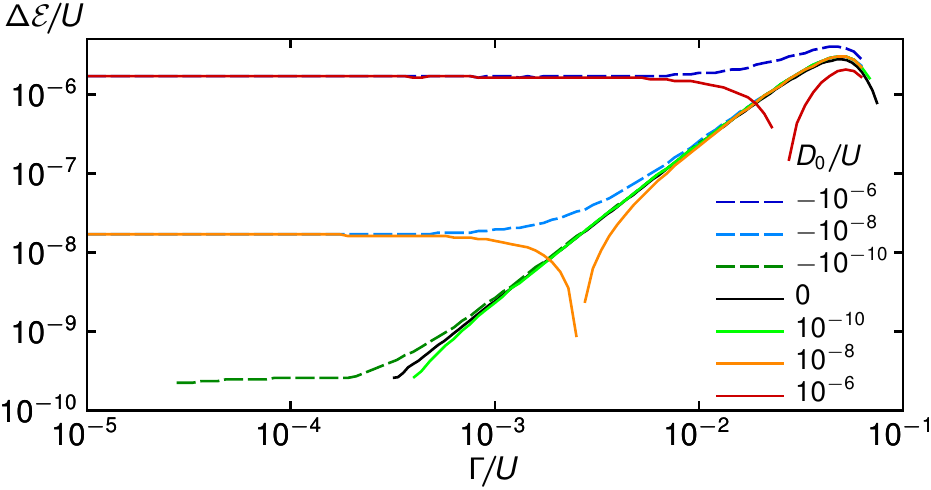}
\caption{
	Generalization of \EFig{fig4}(b) for \mbox{$\GS=0$} to different values of intrinsic magnetic anisotropy of the molecule, $D_0$.
 	All other parameters as in \EFig{fig4}.
	 }
\label{fig:dip-D}
\end{figure}
%++++++++++++++++++++++++++++++++++++++++++++++++++++++++++++++++++++++

In principle, the total anisotropy~$D$ can be predicted assuming that the dependence of the quadrupolar spintronic  exchange field~$\Ds$ on the intrinsic magnetic anisotropy $D_0$ is negligible, \ie, the induced anisotropy does not depend on the intrinsic one.
Figure~\ref{fig:dip-D} presents how the position of the step in the spectral density of ML corresponding to spin-reversal excitations depends on the tunnel-coupling strength~$\Gamma$.
Clearly, for small $\Gamma$, when \mbox{$\Ds \sim\Gamma^2 \ll D_0$}, the value of \mbox{$|D|=\eb/(2\St -1)$} is in practice entirely determined by~$D_0$. Then, for \mbox{$D_0<0$}, the estimated value of~$D$ increases continuously and practically merges with $D_0=0$ curve, if only \mbox{$|D_0|\ll\eb(D_0\!=\!0)$}. For larger values of~$|D_0|$,  the influence of intrinsic magnetic anisotropy is visible as an additional increase of the energy barrier~$\eb$; see the curve for \mbox{$D_0/U=10^{-6}$} in Fig.~\ref{fig:dip-D}.

The situation is qualitatively different for \mbox{$D_0>0$}, when the intrinsic~($D_0$) and spintronic~($\Ds$) contributions to magnetic anisotropy have the opposite sign.
For small values of~$\Gamma$, that is, if \mbox{$\Gamma<\Gamma^\ast$} with the critical value~$\Gamma^\ast$ defined \via the relation  \mbox{$\Ds(\Gamma^*)=-D_0$}, the behavior of the system is determined mainly by~$D_0$. Then, as discussed above, the ground-state doublet is \mbox{$\ket{\pm1/2,-}$}, and the excitation energy~$\eb$ corresponds to the leading transitions to the excited doublet \mbox{$\ket{\pm3/2,-}$}. 
As the tunnel coupling~$\Gamma$ grows and approaches its critical value~$\Gamma^\ast$, this excitation energy gets diminished, and then completely suppressed at \mbox{$\Gamma=\Gamma^*$}, which means that the molecule becomes effectively spin-isotropic, with all states \mbox{$\ket{\Stz,-}$} (for \mbox{$\Stz=-\Sc-1/2, -\Sc+1/2,\ldots,\Sc-1/2,\Sc+1/2$}) being degenerate at this point.
Then, for even larger~$\Gamma$, \mbox{$\Gamma>\Gamma^\ast$}, the spintronic contribution~$\Ds$ starts dominating over the intrinsic component~$D_0$ of magnetic anisotropy, and the spin doublet \mbox{$\ket{\pm(\Sc+1/2),-}$} (or to be precise, \mbox{$\ket{\pm3/2,-}$} for \mbox{$\Sc=1$}) becomes restored as the ground state, with the energy barrier~$\eb$ growing quadratically with~$\Gamma$.

%----------------------------------------------------------------------
\subsection{Superconducting proximity effect (\mbox{$\GS\neq0$}) for finite $D_0$}
\label{sec:D_GS}
%----------------------------------------------------------------------

%++++++++++++++++++++++++++++++++++++++++++++++++++++++++++++++++++++++
\begin{figure}[tb]
\includegraphics[width=0.99\columnwidth]{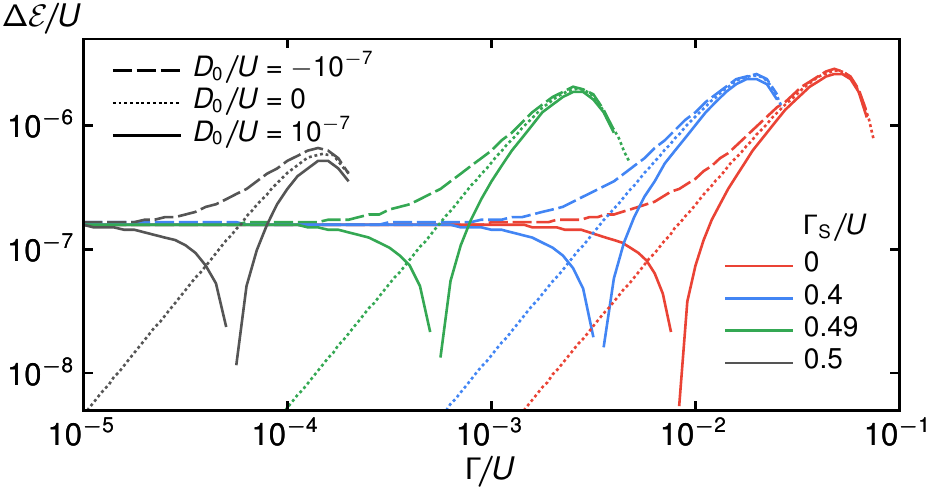}
\caption{
	Analogous to \EFig{fig4}(b), but now plotted also for selected values of the intrinsic magnetic anisotropy constant~$D_0$.
	Note that fine-dashed lines serve as reference lines,
	since they correspond to the results for \mbox{$D_0=0$} shown in \EFig{fig4}(b).
 	All other parameters as in \EFig{fig4}.
		 }
\label{fig:D0-sfA}
\end{figure}
%++++++++++++++++++++++++++++++++++++++++++++++++++++++++++++++++++++++

The superconducting proximity effect for finite $D_0$ can be well understood by assuming that also for 
finite $\GS$ the induced anisotropy $\Ds$ is independent of $D_0$.
In particular, the shift of the anisotropic regime towards smaller values of the tunnel-coupling~$\Gamma$, visible in \EFig{fig2}(b), still takes place for finite $D_0$, retaining the properties discussed in \Sec{D_GS0}.
Similar behavior is also reflected in the spectral properties, which is illustrated 
in \Fig{D0-sfA}.
As expected, for all considered values of~$\GS$, the position of the characteristic step in the ML spectral function remains fixed for \mbox{$D_0<0$} in the weak tunnel-coupling limit and tends to shift only when $\Gamma$ is in the intermediate regime, where \mbox{$|\Ds|\gtrsim|D_0|$}.
For \mbox{$D_0>0$} the situation is similar, although for some critical value of the tunnel-coupling strength~\mbox{$\Gamma=\Gamma^*$} one gets \mbox{$\Ds = -D_0$}, so that the magnetic anisotropy is effectively removed  and the step in the spectral function vanishes. 
Moreover, it is clear that due to the enhancement of $\Ds$ by orders of magnitude
by the superconducting proximity effect (\mbox{$\GS\neq0$}), the condition \mbox{$\Ds=-D_0$}
can be fulfilled already for much smaller values of~$\Gamma$. 
This corroborates the conclusion, that the induced spintronically magnetic anisotropy
$\Ds$ is in essence independent of $D_0$ and the analysis of the spintronic contribution
performed in the previous sections remains valid also for spin-anisotropic molecules,
with the values of the total magnetic anisotropy $D$ being shifted by $D_0$.

%%%%%%%%%%%%%%%%%%%%%%%
\section{Conclusions}
\label{sec:conclusions}
%%%%%%%%%%%%%%%%%%%%%%%

In this paper we have analyzed the interplay of the ferromagnetic and superconducting
contact-generated proximity effects on the transport properties of large-spin molecules.
Employing the density-matrix numerical renormalization group method
allowed us to obtain accurate predictions for the transport behavior
of the considered molecular junction and, in particular,
for the spintronically-induced magnetic anisotropy.
We have shown that the proximity of superconductor can greatly
increase the spintronic anisotropy in the regime of weak and intermediate tunnel coupling to the normal leads.
This effect is particularly strong in the vicinity of the parity-changing transition
of the superconductor-proximized molecule, resulting in a giant 
(by several orders of magnitude) enhancement of magnetic anisotropy.

The mechanism responsible for such
spectacular reinforcement of the spintronic anisotropy
is associated with strong renormalization of Coulomb correlations in the molecule
due to the superconducting proximity effect. The spintronic anisotropy
is induced by spin-resolved charge fluctuations between the molecule
and ferromagnetic contacts. Because strong pairing correlations
tend to lower the Coulomb interactions in the molecule,
these charge fluctuations become enhanced, which effectively gives rise to
a large increase of the generated anisotropy.
On the other hand, when the molecule crosses the parity changing transition
as the superconducting pairing potential becomes increased further,
the molecule enters the Yu-Shiba-Rusinov screened phase \cite{YSRscreening}
and the effect of spintronic anisotropy becomes suppressed.

In addition, we have analyzed the spectral features of the system,
which can give access to detailed quantitative
information about the induced anisotropy.
We have shown that its presence is revealed as a step 
in the local density of states of the molecule, which provides
means for experimental exploration of the discussed phenomena.
It is also important to note that the signatures of the induced spintronic anisotropy
are robust against slight changes of the parameters.
In particular, it remains crucial that the molecular level is at the particle-hole symmetry point,
however, slight detuning from this point, especially 
when the superconducting pairing correlations are relatively large,
should not alter the discussed phenomena.

Our analysis sheds new light on the behavior of hybrid molecular junctions,
especially, on the interplay of the superconducting and ferromagnetic proximity effects
and its signatures in transport properties.
We believe that, by providing new means for manipulating properties of large-spin molecules,
this work shall foster further theoretical and experimental efforts
to investigate unexplored spintronic aspects of the proximity-induced pairing.

\begin{acknowledgments}
The work was supported by National Science Centre in Poland through grants no. DEC-2013/10/E/ST3/00213
and DEC-2017/27/B/ST3/00621.
M.M. also acknowledges financial support from the Knut and Alice Wallenberg Foundation.
K.P.W. acknowledges the support of Alexander von Humboldt Foundation.
\end{acknowledgments}

%%%%%%%%%%%%%%%%%%%%%%%%%%%%%%%%%%%%%%%%%%%%%%%%%%%%%%%%%%%%%%%%%%%
%%%%%%%%%%%%%%%%%%%%%%%%%%%%%%%%%%%%%%%%%%%%%%%%%%%%%%%%%%%%%%%%%%%
\appendix
%%%%%%%%%%%%%%%%%%%%%%%%%%%%%%%%%%%%%%%%%%%%%%%%%%%%%%%%%%%%%%%%%%%
%%%%%%%%%%%%%%%%%%%%%%%%%%%%%%%%%%%%%%%%%%%%%%%%%%%%%%%%%%%%%%%%%%%

%======================================================================
\section{\\ Superconductor-proximized molecule spectrum}
\label{sec:SCmol}
%======================================================================

In this appendix we analyze the eigenstates and eigenvalues of a molecule proximized by a superconducting electrode, \mbox{$\GS>0$}, yet decoupled from ferromagnetic leads, \mbox{$\Gamma=0$}. 
Furthermore, we consider here a more general form of the Hamiltonian~$\Hammol$ for a superconductor-proximized molecule, \EEq{H_SMM}, that involves also the effect of intrinsic uniaxial magnetic anisotropy of the molecule.  This effect is captured by adding a new term~\mbox{$D_0\opQzz$} to~$\Hammol$, with \mbox{$D_0$} standing for the anisotropy constant and $\opQzz$ defined in \Eq{Qzz_def}.
Recall from the main text that the total spin of the molecule, \mbox{$\vopSt=\vopSc+\vopSml$}, is composed of the core spin~$\vopSc$ and the spin~$\vopSml$ of an electron residing in the molecular level (ML). As a result, the molecular Hamiltonian reads now as
\beq\label{Hmol}
	\Hammol^\prime	
	= 
	\Hammol
	+ 
	D_0\opStz^2 
	,
\eeq
with $\Hammol$ given by Eq.~(\ref{eq:H_SMM}).
Note that for conceptual simplicity of the further discussion, in the equation above we have dropped the constant term \mbox{$C=-D_0\St(\St + 1)/3$}. In fact, the value of $C$ is a function of the total spin of 
the molecule, which depends also on its charge state.
However, as long as only the two charge states of lowest energies are relevant for the physics
of the system and the spin of the ground state is well-defined, \ie at $T\ll J,U$, the only consequence 
of $C$ dependence on $\St$ is a small renormalization of $U$ (of the order of $D_0$). 
Thus, it seems justified to neglect the constant term under discussion.

The additional anisotropy term in \Eq{Hmol} will prove helpful in qualitative understanding of the physics 
behind the transition between two different ground states which the molecule undergoes at some critical 
value~$\GSc$. As was shown in Sec.~\ref{sec:D}, its inclusion does not affect qualitatively 
the key results discussed in the main text.

%----------------------------------------------------------------------
\subsection{Exact eigenstates and eigenvalues}
%----------------------------------------------------------------------

First of all, note that, since ferromagnetic leads are
assumed here to be decoupled from the molecule,
the spintronic anisotropy does not occur in the following description.

Despite its complex form, the Hamiltonian~(\ref{Hmol}) can still be solved exactly. For this purpose, let us introduce the suitable basis of the Hilbert space consisting of product states \mbox{$\ket{\core}\otimes\ket{\ML}$}. Here, the state of magnetic core is determined by the $z$-component of its spin~$\vopSc$, \mbox{$\ket{\core}\equiv\ket{\Scz}$}. On the other hand, the state of the molecular orbital (ML) is described by its occupation, that is, \mbox{$\ket{\ML}\in\{0,\up,\down,2\}$}, with 0~(2) representing the unoccupied~(doubly occupied) orbital and \mbox{$\s=\up,\down$} corresponding to the case of the orbital being occupied by a single electron with spin~$\s$.
Next, it is easy to observe that the $z$-component of the total spin, $\Stz$, and the parity of the ML occupation,~$\alpha$ (defined as \mbox{$\alpha=1$} for even- and \mbox{$\alpha=-1$} for odd-parity states), are conserved quantities, and thus, good quantum numbers to be used for labeling the eigenstates. Moreover, the states with different~$\alpha$ are also always characterized by different~$\Stz$ (being integer for one parity value and half-integer for the other).
In consequence, the parity index~$\alpha$ can be omitted in the eigenstate label, so that the eigenstates of Hamiltonian~(\ref{Hmol}) are denoted as $\ket{\Stz,r}$, with $r$ indexing states within a given $\Stz$~subspace. For \mbox{$J>0$}, the possible values of $\Stz$ for the ground-state spin multiplet are in the range from \mbox{$-\Sc-1/2$} to \mbox{$\Sc+1/2$}. As long as \mbox{$|\Stz|\leqslant\Sc$}, all the $\Stz$~subspaces are two-dimensional. Below we first obtain the egienenergies and eigenstates for the situation of the even occupation of ML,
and next we proceed to the case of the ML occupied by a single electron.

\paragraph{The \underline{even} occupation of ML\hspace*{-5pt}}%
(\ie for \mbox{$\ket{\ML}\in \{0,2\}$} and \mbox{$\Stz=\Scz$})---%
In such a case, the ML spin is \mbox{$s=0$} and the term proportional to~$J$ in~\Eq{Hmol} does not contribute. It basically means that states with different values of~$\Stz$ are not mixed, and the two-dimensional Hamiltonian matrix for a given~$\Stz$ can be easily diagonalized, which yields the egienenergies
\begin{subequations}\label{energies_oo}
\begin{gather}\label{energies_ee_a}
	E_{\Stz,\sA}  = \e + U/2 + D_0\Stz^2 + \sqrt{\GS^2 + (U/2+\e)^2} 	,
\\\label{energies_ee_b}
	E_{\Stz,\sB} = \e + U/2 + D_0\Stz^2 - \sqrt{\GS^2 + (U/2+\e)^2} ,
\end{gather}
\end{subequations}
and the corresponding eigenstates
\begin{subequations}\label{states_oo}
\begin{align}\label{states_oo_a}
	&\ket{\Stz,\sA} 	=
		\sqrt{\frac{1-\gamma_\even}{2}} \, \ket{\Stz}\!\otimes\!\ket{0}
\nonumber\\	
	&\hspace*{90pt}	-\sqrt{\frac{1+\gamma_\even}{2}} \, \ket{\Stz}\!\otimes\!\ket{2}	 	
		,
\\\label{states_oo_b}
	&\ket{\Stz,\sB} 	= 
		\sqrt{\frac{1+\gamma_\even}{2}} \, \ket{\Stz}\!\otimes\!\ket{0} 
\nonumber\\		
	&\hspace*{90pt}	+\sqrt{\frac{1-\gamma_\even}{2}} \, \ket{\Stz}\!\otimes\!\ket{2}	 	,
\end{align}
\end{subequations}
with the coefficient~$\gamma_\even$ defined as
\beq
	\gamma_\even
	=
	\frac{\e+U/2}{\sqrt{\Gamma_S^2+(\e+U/2)^2\rule{0pt}{8.5pt}}}.
\eeq 

\paragraph{The \underline{odd} occupation of ML\hspace*{-5pt}}%
(\ie for \mbox{$\ket{\ML}\in \{\up,\down\}$} and \mbox{$|\Stz-\Scz|=1/2$})---%
First of all, a pair of states corresponding to maximal value of \mbox{$|\Stz|=\Sc+1/2$}, namely,
\beq\label{states_maxSz}
	\ket{\Stz=\pm(\Sc+1/2),-}
	=
	\ket{\pm\Sc}
	\otimes
	\ket{\up\!(\down)}
\eeq
are trivial eigenstates with eigenenergies
\beq\label{energies_maxSz}
	E_{\Stz=\pm(\Sc+1/2),-} 
	= 
	\e
	- 
	J\Sc/2
	+
	D_0\Stz^2
	.
\eeq
The reason to take the auxiliary index \mbox{$r=-$} will become clear towards the end of this section. 
The solution of~$\Hammol$ in the remaining subspaces reads
\begin{subequations}\label{energies_ee}
\begin{align}\label{energies_ee_+}
	&E_{\Stz,+}
	=
	\e
	+
	J(\Sc+1)/2
	+
	D_0\Stz^2
	,
\\\label{energies_ee_-}
	&E_{\Stz,-}
	=
	\e
	+
	J\Sc/2
	+
	D_0\Stz^2
	,
\end{align}
\end{subequations}
for eigenenergies, and 
\begin{align}
\label{states_ee}
	\ket{\Stz,\pm} 
	= 
	\frac{1}{\sqrt{2}} 
	\Big(
	&
	\sqrt{1\mp\gamma_\odd(\Stz)}
	\,
	\ket{\Stz-1/2}
	\otimes
	\ket{\up} 
\nonumber\\	
	\mp
	&
	\sqrt{1\pm\gamma_\odd(\Stz)} 
	\,
	\ket{\Stz+1/2}
	\otimes
	\ket{\down} 
	\!
	\Big) 
	,
\end{align}
for eigenstates, with the coefficient~$\gamma_\odd(\Stz)$ given by
\beq
	\gamma_\odd(\Stz)
	=
	\frac{\Stz}{\Sc+1/2}
	.
\eeq
We note that the states denoted in~\Eq{states_ee} as \mbox{$\ket{\Stz,-}$} correspond to components of the spin multiplet \mbox{$\St=\Sc+1/2$}, while those represented as \mbox{$\ket{\Stz,+}$} constitute the spin multiplet \mbox{$\St=\Sc-1/2$}.
Moreover, one can notice that~\Eq{energies_maxSz} is formally equivalent to \Eq{energies_ee_-}---justifying our choice of \mbox{$r=-$} in Eqs.~(\ref{states_maxSz})-(\ref{energies_maxSz})---which is expected because the spin multiplet \mbox{$\St=\Sc+1/2$} should be split only \via the term proportional to~$D_0$.
We also remind that eigenenergies~$E_{\Stz,-}$ and~$E_{\Stz,+}$, \Eq{energies_ee}, should be understood as the ground-state energies for the spin multiplet \mbox{$\St=\Sc+1/2$} (if \mbox{$J > 0$}) and \mbox{$\St=\Sc-1/2$} (if \mbox{$J < 0$}), respectively.
In the limit of non-magnetic leads (\mbox{$p = 0$}) and for \mbox{$D_0 = 0$}, as considered in \EFig{fig2}(a) in the main text, the $SU(2)$ spin symmetry is recovered and the states in all respective spin multiplets are degenerate.

Finally, let us point out that the model of the molecule used in this work, and represented by the Hamiltonian~$\Hammol$, \Eq{Hmol}, differs slightly from that used by Misiorny~\etal in Ref.~\cite{Misiorny2013}. The difference concerns the way how the \emph{intrinsic} uniaxial magnetic anisotropy is included into the model. Specifically, here we assume that the \emph{total} spin~$\vopSt$ of the molecule is subject to magnetic anisotropy, while in Ref.~\cite{Misiorny2013} only the \emph{core} spin~$\vopSc$ is affected by magnetic anisotropy.

%----------------------------------------------------------------------
\subsection{Andreev bound states\label{app:ABS}}
%----------------------------------------------------------------------

%++++++++++++++++++++++++++++++++++++++++++++++++++++++++++++++++++++++
\begin{figure}[tb]
\includegraphics[width=0.99\columnwidth]{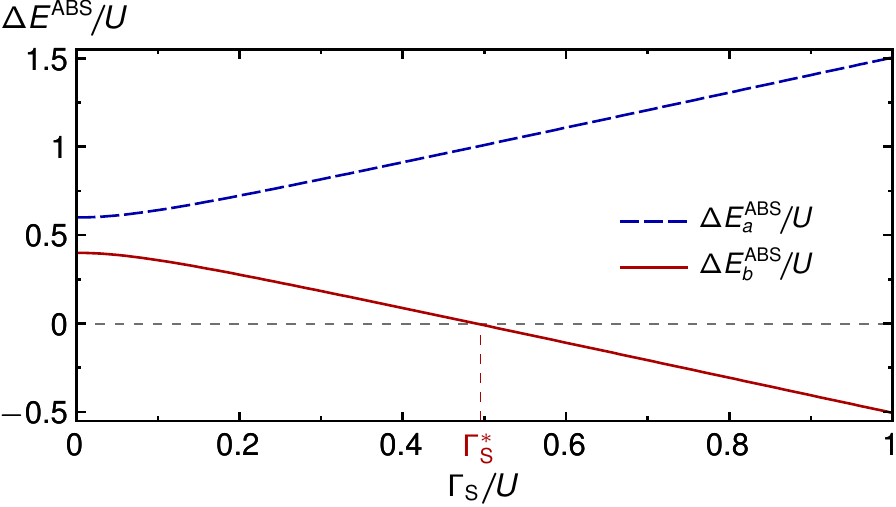}
\caption{
	The energies of the Andreev bound states~$\Delta E_{\sA(\sB)}^\text{ABS}$ [see \Eq{ABS_F}] presented as a function of the coupling strength~$\GS$ of the molecule to the superconductor in the case of the ferromagnetic exchange coupling between spins of the ML and the magnetic core of the molecule.
	Parameters used: \mbox{$\Sc=1$}, \mbox{$\delta/U=0.1$}, \mbox{$J/U=10^{-3}$} and \mbox{$D_0=0$}.
	 }
\label{fig:ABS}
\end{figure}
%++++++++++++++++++++++++++++++++++++++++++++++++++++++++++++++++++++++

Let us begin with the case of \mbox{$\GS=0$} and the physically reasonable limit of \mbox{$|D_0|\ll|J|\ll U$}. Then, for \mbox{$|\e+U/2|\ll U$}, the ground state of the molecule corresponds to the ML occupied by a single electron. 
As in the main text, we focus on the situation when the exchange coupling~$J$ arising between the spin of such an electron and the core spin of the molecule is ferromagnetic (\mbox{$J > 0$}), which means that the ground spin multiplet is characterized by spin \mbox{$\St=\Sc+1/2$}. The relevant excited charge multiplets correspond to empty or doubly occupied ML, with excitation energies of the order of \mbox{$U/2\pm\delta$}, where \mbox{$\delta=\e+U/2$} is the detuning from the particle-hole symmetry point---note that \mbox{$\delta=0$} in the main text.

The situation becomes more complex for \mbox{$\GS>0$}, as in such a case the empty and doubly occupied states of ML are hybridized due to the proximity of superconductor, see \Eq{states_oo}. Nevertheless, the two charge multiplets with the even occupation of ML remain relevant excitations, forming the so-called \emph{Andreev bound states}. Specifically, at low temperature (\mbox{$T\ll2\Sc|D_0|, J$}), when only the ground-state spin doublet \mbox{$\ket{\pm(\Sc+1/2),-}$} is occupied, energies of relevant excitations are defined as
\begin{align}\label{ABS_F_def}
	\Delta E_{\sA(\sB)}^\text{ABS} 
	\equiv\ & 
	E_{\Sc,\sA(\sB)}
	-
	E_{\Sc+1/2,-}
\nonumber\\
	=\ &
	E_{-\Sc,\sA(\sB)}
	-
	E_{-(\Sc+1/2),-}
	,
\end{align}
which gives
\begin{subequations}\label{ABS_F}
\begin{gather}\label{ABS_F_a}
\Delta E_{\sA}^\text{ABS} 
	= \Delta E	+\sqrt{\GS^2+\delta^2}
%	\qquad
%	{\scriptstyle (J>0)}
,
\\ \label{ABS_F_b}
\Delta E_{\sB}^\text{ABS} 
	= \Delta E	-\sqrt{\GS^2+\delta^2}
%	\qquad
%	{\scriptstyle (J>0)}
.
\end{gather}
\end{subequations}
with
\begin{equation}
	\Delta E	
	=
	U/2 + 	J\Sc/2 + |D_0|(\Sc+1/4)
	.
\end{equation}
Note that there exist also other Andreev bound states, involving transitions between different excited
states \cite{Pawlicki2018Aug}. However, they are insignificant for the transport properties analyzed here and in the main article, 
so that we do not discuss them in detail.

In \Fig{ABS} we show how the energies of the Andreev bound states evolve as a function of the coupling~$\GS$ for \mbox{$\delta>0$}.
For \mbox{$\GS=0$}, the two energies correspond to the excitation energies to empty ($\Delta E_{\sB}^\text{ABS}$) or doubly occupied ($\Delta E_{\sA}^\text{ABS}$) states of ML, which are of the order of~$U/2$, split by particle-hole asymmetry~$\delta$.
Additional spin-related contribution to the splitting occurs also due to the interaction of the ML spin with the magnetic core of the molecule, $J\Sc/2$, as well as due to magnetic anisotropy, \mbox{$|D_0|(\Sc + 1/4)$}. In the proximity of superconductor, \mbox{$\GS>0$}, the two channels of excitations no longer correspond to particle or hole processes.
On the contrary, two Andreev bound states are formed, whose energies develop continuously from the values observed at \mbox{$\GS=0$}, always increasing the splitting as~$\GS$ grows. Note that the spin-related contribution to the splitting of the two channels does not depend on~$\GS$.

Importantly, one can observe in \Fig{ABS} that the excitation energy associated with the transition denoted by `b' becomes equal to 0 when~$\GS$ is tuned to some critical value \mbox{$\GS=\GSc$},
\beq\label{GSc_F}
	\GSc
	=
	\sqrt{
	\Delta E^2
	-
	\delta^2
	}
	.
\eeq
In fact, $\GSc$ is the quantum phase transition point, as the negative excitation energy essentially means the change in the ground state---with the new phase for \mbox{$\GS>\GSc$} being the Yu-Shiba-Rusinov screened (evenly-occupied ML) phase~\cite{YSRscreening}.

In the case considered in the main text, that is, for an intrinsically spin-isotropic molecule (\mbox{$D_0 = 0$}) tuned to the particle-hole symmetry point (\mbox{$\delta = 0$}) and attached to ferromagnetic leads, the anisotropy constant~$D_0$ in \Eq{GSc_F} should be replaced by the quadrupolar spintronic field~$D$.
However, for parameters used there (\mbox{$J/U = 10^{-3}$} and \mbox{$\Sc = 1$}), the contribution to~$\GSc$ due to spintronic anisotropy is of the order of \mbox{$|D|/U\lesssim10^{-6}$}, and thus, it can be neglected. Consequently, the estimated value of the relevant quantum phase transition point is
\beq\label{GSc_approx}
	\GSc/U
	\approx
	(1+J\Sc/U)/2
	=
	0.5005
	.
\eeq

Finally, to complete this discussion, we note that an analogous analysis can be conducted also for the case of the antiferromagnetic exchange coupling between the ML and core spin, \mbox{$J < 0$}.
Then, the key difference with respect to the case of \mbox{$J > 0$} discussed above is that the state acting as the ground-state spin doublet for a singly occupied ML is \mbox{$\ket{\pm(\Sc-1/2),+}$}. As a result, the excitation energies associated with Andreev bound states are [we introduce tilde to distinguish the $J<0$ case
from the $J>0$ case in Eqs.~(\ref{ABS_F_def})-(\ref{GSc_approx})]
\begin{align}\label{ABS_AF_def}
	\Delta\widetilde{E}_{\sA(\sB)}^\text{ABS} 
	\equiv\ &
	E_{\Sc,\sA(\sB)}
	-
	E_{\Sc-1/2,+}
\nonumber\\
	=\ &
	E_{-\Sc,\sA(\sB)}
	-
	E_{-(\Sc-1/2),+},
\end{align}
which gives 
\begin{subequations}\label{ABS_AF}
\begin{gather}\label{ABS_AF_a}
	\Delta\widetilde{E}_{\sA}^\text{ABS} =
	 \Delta\widetilde{E} + \sqrt{\GS^2+\delta^2}
,
\\ \label{ABS_AF_b}
	\Delta\widetilde{ E}_{\sB}^\text{ABS} =
	\Delta\widetilde{E} - \sqrt{\GS^2+\delta^2}
,
\end{gather}
\end{subequations}
with 
\begin{equation}
	\Delta\widetilde{E}
	=
	U/2 + |J|\Sc/2 - |D_0|(\Sc-1/4)
	.
\end{equation}
Consequently, the critical value of~$\widetilde{\Gamma}_\text{S}^\ast$ takes the following form
\beq\label{GSc_AF}
	\widetilde{\Gamma}_\text{S}^\ast
	=
	\sqrt{
	\Delta\widetilde{E}^2
	-
	\delta^2
	}
	.
\eeq
%

%======================================================================
\section{Discussion of the spectral function properties in full range of energies}
\label{sec:sfA}
%======================================================================

%++++++++++++++++++++++++++++++++++++++++++++++++++++++++++++++++++++++
\begin{figure}[tb]
\centering
\includegraphics[width=0.99\columnwidth]{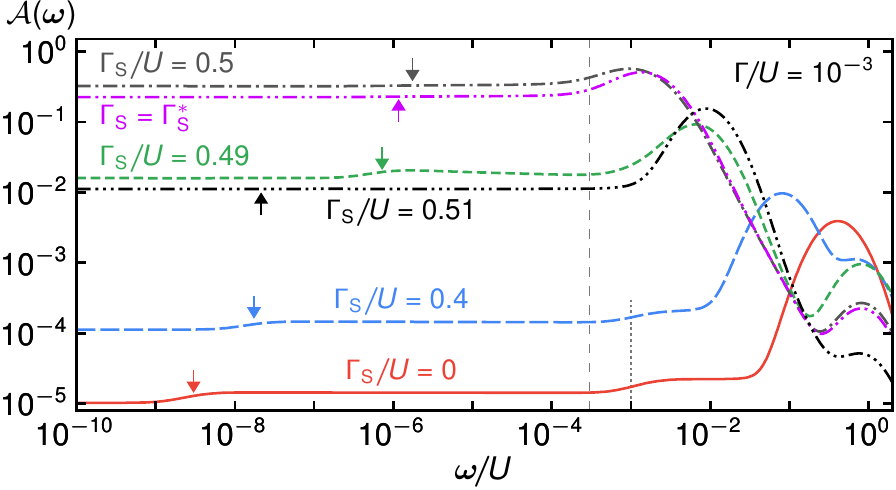}
\caption{
	Normalized spectral function of the ML, $\sfA(\w)$, corresponding to \EFig{fig4}(a) but now plotted without scaling to~\mbox{$\sfA_0\equiv\sfA(\w=0)$} and in the full range of energies~$\w$. 
	Arrows indicate the expected positions of the step related to~$\eb$ obtained from the compensation method explained in Appendix~\ref{sec:Qzz0}. 
	Note that, unlike in \EFig{fig4}(a), here~$\sfA(\w)$ is shown on the logarithmic scale. 
	The vertical thin dashed line indicates the maximal range of energies~$\w$ presented in \EFig{fig4}(a).
	 }
\label{fig:sfA}
\end{figure}
%++++++++++++++++++++++++++++++++++++++++++++++++++++++++++++++++++++++

Figure~\ref{fig:sfA} presents the spectral function of the ML, $\sfA(\w)$, in the full range of energies~$\w$. Analogously as \EFig{fig4}(a) in the main text, \Fig{sfA} illustrates the evolution of the spectra for different~$\GS$. 
However, in order to resolve better the high-energy features in the spectral function, here $\sfA(\w)$ is plotted on the logarithmic scale and without rescaling by \mbox{$\sfA_0\equiv\sfA(\w=0)$}. We also show an additional curve for \mbox{$\GS/U = 0.51$}, which is not considered in \EFig{fig4}.

The first observation is that~$\sfA_0$ is very small for \mbox{$\GS = 0$}, with \mbox{$\sfA_0/U\approx 10^{-5}$}. This effect stems from the fact that for the tunnel coupling under consideration, \mbox{$\Gamma/U = 10^{-3}$}, temperature used in calculations, \mbox{$T/U = 10^{-10}$}, is still larger than the Kondo temperature, \mbox{$T > \TK$}. Thus, the results obtained for such parameters at the particle-hole symmetry point (\mbox{$\delta=0$}) actually correspond to the Coulomb blockade regime. Since the conductance~$G$ is determined there by electron cotunneling processes and it scales as \mbox{$G(\Gamma) \propto \Gamma^2$}, it is small in the limit of the weak tunnel coupling (i.e., when~$\Gamma$ is small). 
On the other hand, the conductance increases when the cotunneling processes are enhanced, which takes place when the excitation energies to the states with the even occupation of the ML diminish---as it happens when increasing~$\GS$; see Appendix~\ref{sec:SCmol}.

As explained in the main text, there appears a step in the spectral function localized at \mbox{$\omega\approx\eb=(2\St-1)|D|$}, whose position becomes shifted towards higher energies as~$\GS$ gets augmented. 
However, at the same time, the relative hight of the step becomes diminished to such an extent that in the vicinity of the critical point, \mbox{$\GS\approx\GSc$}, one cannot distinguish the step any longer; see the curve for \mbox{$\GS/U = 0.5$} in \Fig{sfA}. 
Positions of the steps for different~$\GS$ agree with the values of~$\eb$ estimated from the compensation method described in Appendix~\ref{sec:Qzz0}, and they are indicated in \Fig{sfA} by the arrows. Note that even though the compensation method allows for calculating~$\eb$ for arbitrary values of~$\GS$, no particular spectral features appear in~$\sfA(\w)$ for \mbox{$\GS\gtrsim\GSc$}, \ie, close to or within the Yu-Shiba-Rusinov screened phase. For this reason, we did not present in \EFig{fig4} the results for \mbox{$\GS\geqslant\GSc$}.

Furthermore, in the limit of \mbox{$\GS\ll\GSc$}, the spectral function exhibits also a second step at energies \mbox{$|\w|\approx J$}, marked by the finely dashed vertical line in \Fig{sfA}. This step arises naturally in the present model due to the ferromagnetic exchange coupling~$J$, and it corresponds to transitions between the ground-state doublet \mbox{$\ket{\pm(\Sc + 1/2), -}$} (belonging to the spin multiplet \mbox{$\St = \Sc + 1/2$}) and the doublet \mbox{$\ket{\pm(\Sc - 1/2), +}$} of the excited multiplet with spin \mbox{$\St = \Sc - 1/2$}. 
For \mbox{$\GS\gtrsim\GSc$}, this step becomes suppressed  by charge fluctuations, which become very efficient close to the critical point~$\GSc$, as explained in the following.

For \mbox{$\GS=0$}, charge fluctuations require very high energies, \mbox{$\w\approx U/2$}, and correspond to the so-called Hubbard peak in the spectral density $\sfA(\w)$. 
However, due to the proximity of superconductor, \mbox{$\GS>0$}, the excitation energy associated with charge fluctuations is decreased, see \Eq{ABS_F_b}, until it reaches~0 at \mbox{$\GS=\GSc$}, and next increases again after the new ground state has been established. 
This effect is visible in \Fig{sfA}, where the position of the Hubbard peak becomes shifted towards smaller energies with~$\GS$ approaching the critical point as \mbox{$\w \approx \Delta E_{\sB}^{\text{ABS}}(\GS)$}, and it gets shifted back towards larger~$\w$ for \mbox{$\GS>\GSc$}, as expected from \Eq{ABS_F_b}. 
Note, however, that due to the finite tunnel coupling~$\Gamma$, the critical point is slightly shifted from $\GS=\GSc$ to a value smaller by \mbox{$\sim\Gamma$}, \cf \EFig{fig3}. This fact is the reason why the Hubbard peak for \mbox{$\GS=\GSc$ } in \Fig{sfA} is not positioned at~\mbox{$\w=0$}, but approximately at \mbox{$\w/U=\Gamma/U=10^{-3}$}. 
At the same time, the excitation energy in the second excitation channel follows \mbox{$\w \approx |\Delta E_{\sA}^{\text{ABS}}(\GS)|$}, increasing with $\GS$ according to \Eq{ABS_F_a}. This can be observed in \Fig{sfA} as the second Hubbard peak occurring for \mbox{$\GS>0$} at \mbox{$\w/U>1/2$}.

%======================================================================
\section{Compensation method for calculation of induced anisotropy}
\label{sec:Qzz0}
%======================================================================

As mentioned in the introductory section of the main text, the effective Hamiltonian describing magnetic molecule has a general form
\beq
	\Hameff 
	= 
	B \opStz 
	+ 
	D \opQzz
\eeq
with~$B$ and~$D$ representing magnetic field and magnetic anisotropy, respectively. 
While assuming that \mbox{$B=0$}, we focus here on analyzing the role played by finite~$D$, which can be either of intrinsic (due to spin-orbit interaction~\cite{Kahn_book,Boca_book,Gatteschi_book}) or extrinsic (\ie, generated spintronically~\cite{Misiorny2013}) origin.
	As a result, one can in general think of~$D$ as composed of two contributions, \mbox{$D=D_0+\Ds$}, with $D_0$ representing the intrinsic contribution [introduced in \Eq{Hmol}] and $\Ds$ standing for the spintronic contribution, which arises due to the proximity of ferromagnets. Note that in the main text we consider mostly the case of an intrinsically spin-isotropic molecule (\mbox{$D_0=0$}), so that \mbox{$D=\Ds$}.
In real systems, $D_0$ is a property of a particular type of molecule and it can be either positive or negative, whereas \mbox{$\Ds<0$}, since the second-order electron tunneling processes always decrease the energy of the ground state~\cite{Misiorny2013}.

Importantly, in a system tuned to the particle-hole symmetry point (\mbox{$\delta=0$})---to avoid the occurrence of the spintronic dipolar field---$\Ds$ can be adjusted in a limited range by changing the tunnel coupling to ferromagnetic leads or by varying the spin polarization of the leads. Thus, one can imagine that in a molecule with small `easy-plane' magnetic anisotropy (\mbox{$D_0>0$}) that is attached to ferromagnets, it is in principle possible to compensate~$D_0$ by tuning~$\Ds$, so that the molecule becomes effectively spin-isotropic.
This situation corresponds to exact vanishing of the spin-quadrupole moment, \Eq{Qzz_def}, at all temperatures~$T$, \mbox{$\avQzz(T)=0$}. On the other hand, \mbox{$\avQzz>0$} indicates the `easy-axis' magnetic anisotropy (\mbox{$D<0$}), and \mbox{$\avQzz<0$} is a hallmark of the `easy-plane' magnetic anisotropy (\mbox{$D>0$}).

Actually, based on the discussion above, we come up with a particularly convenient approach allowing for theoretical analysis of the induced magnetic anisotropy~$\Ds$ for a given model. Specifically, we treat  the intrinsic component of magnetic anisotropy---in this context denoted by $-\tD$---as a parameter that can be changed in calculations to obtain $\avQzz(\tD)$. The opposite sign of $\tD$ reflects its \emph{hypothetical} character, in contrast to a clear physical meaning of $D_0$. Then, $\Ds$ is equal to  $\tD$ that satisfies the condition \mbox{$\avQzz(\tD)=0$}.  Within such a  theoretical framework, which in the main text we refer to as the \emph{compensation method}, $\Ds$ has a precise meaning for any set of model parameters, irrespective of its signatures in the local spectral density of ML.

%++++++++++++++++++++++++++++++++++++++++++++++++++++++++++++++++++++++
\begin{figure}[tb]
\includegraphics[width=0.99\columnwidth]{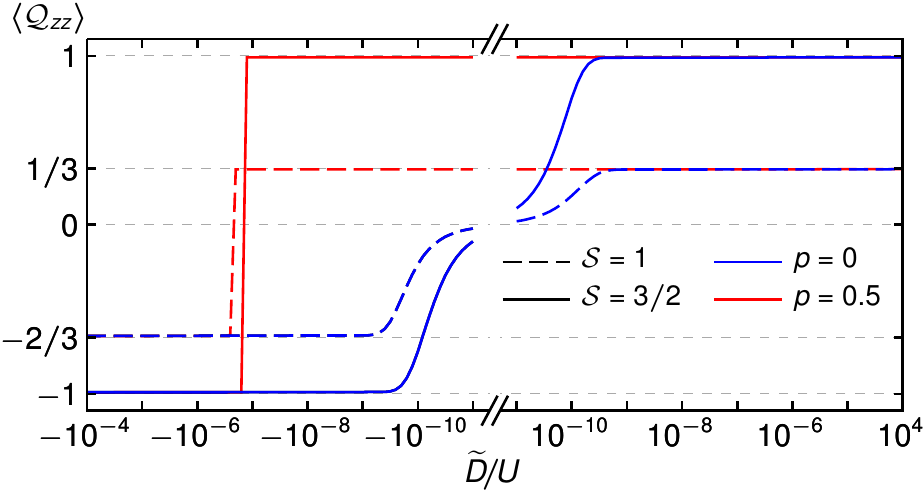}
\caption{
	Illustration of the compensation method for calculating the spintronic component of magnetic anisotropy for  two different examples of the spin magnitude: \mbox{$\St=1$} (dashed lines) and \mbox{$\St=3/2$} (solid lines).
	Moreover, blue lines correspond to the case of non-magnetic (\mbox{$p=0$}) leads, whereas the red ones represent the case of ferromagnetic leads with \mbox{$p=0.5$}.
	Parameters used in calculations: \mbox{$U/W=0.5$}, \mbox{$T/U=10^{-10}$}, \mbox{$J/U=10^{-3}$} and \mbox{$\Gamma/U = 0.01$}.	
	See Appendix~\ref{sec:Qzz0} for further explanation. 
	 }
\label{fig:Qzz}
\end{figure}
%++++++++++++++++++++++++++++++++++++++++++++++++++++++++++++++++++++++

The working principle of the method under discussion is illustrated in \Fig{Qzz}.
To begin with, we first calculate $\avQzz(\tD)$ in the presence of only normal, metallic and non-magnetic leads. This case is plotted in \Fig{Qzz} with blue lines: a solid line for \mbox{$\St=3/2$}, 
as in the main text, and, for comparison,  with a dashed line for \mbox{$\St=1$}.
Clearly, for \mbox{$\tD < 0$} and \mbox{$|\tD| \ll T$}, \ie, for large positive (`easy-plane') hypothetical $D_0$, 
\beq
	\avQzz 
	= 
	(\SzMM^{\rm min})^2
	-
	\frac{\SMM(\SMM+1)}{3} < 0
\eeq 
for \mbox{$\tD < 0$} and \mbox{$|\tD| \ll T$},
where the minimal $z$-component of the molecular spin is \mbox{$\SzMM^{\rm min} = 0$} for integer and \mbox{$\SzMM^{\rm min}=1/2$} for half-integer value of the spin.
At \mbox{$\tD=0$}, the $SU(2)$ spin symmetry guarantees \mbox{$\avQzz=0$}, which is for finite $T$ obtained smoothly with increasing $\tD$. Furthermore, similar smooth increase is obtained for positive~$\tD$, until $\avQzz$ reaches for \mbox{$\tD \gg T$} its maximal value
\beq
	\avQzz=\St^2-\St(\St+1)/3
\eeq
for \mbox{$\tD > 0$} and \mbox{$\tD \ll T$}.

The situation changes substantially when the normal non-magnetic leads become replaced by ferromagnets. Then, the asymptotic values of $\avQzz(\tD)$ remain the same as above, but the zero is shifted to a finite negative value of $\tD$. This shift basically corresponds to the cancellation of the `easy-plane' magnetic anisotropy of a hypothetical molecule for a given set of $p$, $\Gamma$, $J$, and other model parameters. In other words, the molecule becomes spin-isotropic due to the presence of a spintronic component of magnetic anisotropy, which implies that the quadrupolar exchange field~$\Ds$ is of the `easy-axis' type (\ie, \mbox{$\Ds<0$}, as expected), and its magnitude~\mbox{$\Ds=\tD$} can be determined numerically from the condition \mbox{$\avQzz(\tD)=0$}.

It is worth emphasizing that the key advantage of the method for deriving~$\Ds$ described above is that $\Ds$  can also be calculated in this way in the strong-coupling Kondo regime and in the Yu-Shiba-Rusinov phase, where the usual spectral signatures of the magnetic anisotropy are suppressed.
Moreover, a direct comparison with the results obtained from the analysis of the relevant spectral function (see \EFig{fig4}) shows that the compensation method provides in fact a very accurate tool for calculating $\Ds$ also in the regime, where it can be effectively estimated from the ML spectral function. 
On the other hand, the main technical disadvantage of the method is that each point on the $\avQzz(\tD)$-plot corresponds to a separate run of NRG calculation, and for very small $\tD$ one needs to keep very low $T$ and a large number of states at each iteration step in order to obtain reliable results. In practice, this inconvenience can be partially circumvented by  iteratively increasing the sampling of $\avQzz(\tD)$ in the vicinity of the zero, which can significantly reduce the necessary number of points.
Finally, note that a finite physical $D_0$ by definition only shifts the values of $\tD$ by a constant,
therefore, taking it into account does not introduce any additional complications. 

%++++++++++++++++++++++++++++++++++++++++++++++++++++++++++++++++++++++
\begin{figure}[tb]
\includegraphics[width=0.99\columnwidth]{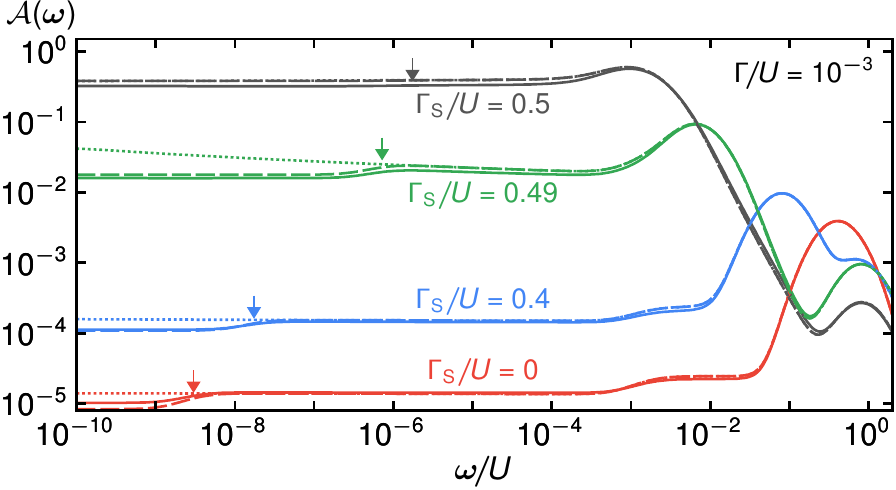}
\caption{
	Comparison of the spectral functions of ML, $\sfA(\w)$, for a finite spintronic component~$\Ds$  to magnetic anisotropy (shown as solid lines with the same color code as in~\Fig{sfA}) with those for \mbox{$p=0$} (and thus, for \mbox{$\Ds=0$}, but with \mbox{$D_0=-\tD$}; dashed lines). The arrows indicate \mbox{$\w=D_0$} for the corresponding curve.
	Dotted lines present the results for the spin-anisotropic case (\mbox{$p=0$} and \mbox{$D_0=0$}). 
	Other parameters as in \Fig{sfA}.
	 }
\label{fig:A0_Ap_AD}
\end{figure}
%++++++++++++++++++++++++++++++++++++++++++++++++++++++++++++++++++++++

In order to prove that the compensation method and the analysis of spectral features in transport indeed yield the same results for the spintronic magnetic anisotropy~$\Ds$, in \Fig{A0_Ap_AD} we directly compare the relevant spectral densities of ML corresponding to three situations: 
\begin{enumerate}[leftmargin=*,label=(\arabic*)]
\item
A spin-isotropic (\mbox{$D_0=0$}) molecule attached to ferromagnetic leads characterized by the spin-polarization coefficient of \mbox{$p=0.5$} (solid lines); \mbox{$D=\Ds$} can be estimated as $\tD$ using the compensation method.
\item 
A spin-anisotropic molecule with \mbox{$D_0 = \tD$} corresponding to the case (1) attached to non-magnetic leads (\ie, for \mbox{$p=0$}), implying that \mbox{$\Ds=0$} and \mbox{$D=D_0$} (dashed lines).
\item 
The case of \mbox{$p=0$} and \mbox{$D_0=0$} for comparison (dotted lines).
\end{enumerate}
One can see that for \mbox{$|\w| \gg D$} the spectral functions~$\sfA(\w)$ corresponding to (2) and (3) are indistinguishable, while for (1) they are slightly reduced for \mbox{$|\w|/U < 1/10$}. 
This is a consequence of the proximity of ferromagnetic lead.

The discussion of the spectral features for \mbox{$|\w| \gg  D$}
was presented in Appendix~\ref{sec:sfA}.
For weak enough $\GS$, curves corresponding to cases (1) and (2) exhibit a step at \mbox{$\w \approx \eb = (2\St-1)|D|$} and a plateau for lower energies---both being characteristic signatures of the presence of magnetic anisotropy. 
In all cases (\ie, also when the step for large $\GS$ is absent) curves for (1) and (2) differ only by a small rescaling factor due to finite $p$ in (1). This is yet another prove that the compensation method for determining~$\Ds$ allows for calculation of the physically relevant quantity and does not lead to non-physical predictions concerning spectra, even in the regimes where the spectral signatures indicating magnetic  anisotropy are difficult to recognize.

%\bibliography{bib_SMM_SC}

\begin{thebibliography}{47}%
\makeatletter
\providecommand \@ifxundefined [1]{%
 \@ifx{#1\undefined}
}%
\providecommand \@ifnum [1]{%
 \ifnum #1\expandafter \@firstoftwo
 \else \expandafter \@secondoftwo
 \fi
}%
\providecommand \@ifx [1]{%
 \ifx #1\expandafter \@firstoftwo
 \else \expandafter \@secondoftwo
 \fi
}%
\providecommand \natexlab [1]{#1}%
\providecommand \enquote  [1]{``#1''}%
\providecommand \bibnamefont  [1]{#1}%
\providecommand \bibfnamefont [1]{#1}%
\providecommand \citenamefont [1]{#1}%
\providecommand \href@noop [0]{\@secondoftwo}%
\providecommand \href [0]{\begingroup \@sanitize@url \@href}%
\providecommand \@href[1]{\@@startlink{#1}\@@href}%
\providecommand \@@href[1]{\endgroup#1\@@endlink}%
\providecommand \@sanitize@url [0]{\catcode `\\12\catcode `\$12\catcode
  `\&12\catcode `\#12\catcode `\^12\catcode `\_12\catcode `\%12\relax}%
\providecommand \@@startlink[1]{}%
\providecommand \@@endlink[0]{}%
\providecommand \url  [0]{\begingroup\@sanitize@url \@url }%
\providecommand \@url [1]{\endgroup\@href {#1}{\urlprefix }}%
\providecommand \urlprefix  [0]{URL }%
\providecommand \Eprint [0]{\href }%
\providecommand \doibase [0]{http://dx.doi.org/}%
\providecommand \selectlanguage [0]{\@gobble}%
\providecommand \bibinfo  [0]{\@secondoftwo}%
\providecommand \bibfield  [0]{\@secondoftwo}%
\providecommand \translation [1]{[#1]}%
\providecommand \BibitemOpen [0]{}%
\providecommand \bibitemStop [0]{}%
\providecommand \bibitemNoStop [0]{.\EOS\space}%
\providecommand \EOS [0]{\spacefactor3000\relax}%
\providecommand \BibitemShut  [1]{\csname bibitem#1\endcsname}%
\let\auto@bib@innerbib\@empty
%</preamble>
\bibitem [{\citenamefont {Jacobson}\ \emph {et~al.}(2015)\citenamefont
  {Jacobson}, \citenamefont {Herden}, \citenamefont {Muenks}, \citenamefont
  {Laskin}, \citenamefont {Brovko}, \citenamefont {Stepanyuk}, \citenamefont
  {Ternes},\ and\ \citenamefont {Kern}}]{Jacobson_Nat.Commun.6/2015}%
  \BibitemOpen
  \bibfield  {author} {\bibinfo {author} {\bibfnamefont {P.}~\bibnamefont
  {Jacobson}}, \bibinfo {author} {\bibfnamefont {T.}~\bibnamefont {Herden}},
  \bibinfo {author} {\bibfnamefont {M.}~\bibnamefont {Muenks}}, \bibinfo
  {author} {\bibfnamefont {G.}~\bibnamefont {Laskin}}, \bibinfo {author}
  {\bibfnamefont {O.}~\bibnamefont {Brovko}}, \bibinfo {author} {\bibfnamefont
  {V.}~\bibnamefont {Stepanyuk}}, \bibinfo {author} {\bibfnamefont
  {M.}~\bibnamefont {Ternes}}, \ and\ \bibinfo {author} {\bibfnamefont
  {K.}~\bibnamefont {Kern}},\ }\bibfield  {title} {\enquote {\bibinfo {title}
  {{Quantum engineering of spin and anisotropy in magnetic molecular
  junctions}},}\ }\href {\doibase 10.1038/ncomms9536} {\bibfield  {journal}
  {\bibinfo  {journal} {Nat. Commun.}\ }\textbf {\bibinfo {volume} {6}},\
  \bibinfo {pages} {8536} (\bibinfo {year} {2015})}\BibitemShut {NoStop}%
\bibitem [{\citenamefont {Donati}\ \emph {et~al.}(2016)\citenamefont {Donati},
  \citenamefont {Rusponi}, \citenamefont {Stepanow}, \citenamefont
  {W{\ifmmode\ddot{a}\else\"{a}\fi}ckerlin}, \citenamefont {Singha},
  \citenamefont {Persichetti}, \citenamefont {Baltic}, \citenamefont {Diller},
  \citenamefont {Patthey}, \citenamefont {Fernandes}, \citenamefont {Dreiser},
  \citenamefont
  {{\ifmmode\check{S}\else\v{S}\fi}ljivan{\ifmmode\check{c}\else\v{c}\fi}anin},
  \citenamefont {Kummer}, \citenamefont {Nistor}, \citenamefont {Gambardella},\
  and\ \citenamefont {Brune}}]{Donati16}%
  \BibitemOpen
  \bibfield  {author} {\bibinfo {author} {\bibfnamefont {F.}~\bibnamefont
  {Donati}}, \bibinfo {author} {\bibfnamefont {S.}~\bibnamefont {Rusponi}},
  \bibinfo {author} {\bibfnamefont {S.}~\bibnamefont {Stepanow}}, \bibinfo
  {author} {\bibfnamefont {C.}~\bibnamefont
  {W{\ifmmode\ddot{a}\else\"{a}\fi}ckerlin}}, \bibinfo {author} {\bibfnamefont
  {A.}~\bibnamefont {Singha}}, \bibinfo {author} {\bibfnamefont
  {L.}~\bibnamefont {Persichetti}}, \bibinfo {author} {\bibfnamefont
  {R.}~\bibnamefont {Baltic}}, \bibinfo {author} {\bibfnamefont
  {K.}~\bibnamefont {Diller}}, \bibinfo {author} {\bibfnamefont
  {F.}~\bibnamefont {Patthey}}, \bibinfo {author} {\bibfnamefont
  {E.}~\bibnamefont {Fernandes}}, \bibinfo {author} {\bibfnamefont
  {J.}~\bibnamefont {Dreiser}}, \bibinfo {author} {\bibfnamefont
  {{\ifmmode\check{Z}\else\v{Z}\fi}.}~\bibnamefont
  {{\ifmmode\check{S}\else\v{S}\fi}ljivan{\ifmmode\check{c}\else\v{c}\fi}anin}},
  \bibinfo {author} {\bibfnamefont {K.}~\bibnamefont {Kummer}}, \bibinfo
  {author} {\bibfnamefont {C.}~\bibnamefont {Nistor}}, \bibinfo {author}
  {\bibfnamefont {P.}~\bibnamefont {Gambardella}}, \ and\ \bibinfo {author}
  {\bibfnamefont {H.}~\bibnamefont {Brune}},\ }\bibfield  {title} {\enquote
  {\bibinfo {title} {{Magnetic remanence in single atoms}},}\ }\href {\doibase
  10.1126/science.aad9898} {\bibfield  {journal} {\bibinfo  {journal}
  {Science}\ }\textbf {\bibinfo {volume} {352}},\ \bibinfo {pages} {318}
  (\bibinfo {year} {2016})}\BibitemShut {NoStop}%
\bibitem [{\citenamefont {Mannini}\ \emph {et~al.}(2009)\citenamefont
  {Mannini}, \citenamefont {Pineider}, \citenamefont {Sainctavit},
  \citenamefont {Danieli}, \citenamefont {Otero}, \citenamefont
  {Sciancalepore}, \citenamefont {Talarico}, \citenamefont {Arrio},
  \citenamefont {Cornia}, \citenamefont {Gatteschi},\ and\ \citenamefont
  {Sessoli}}]{Mannini09}%
  \BibitemOpen
  \bibfield  {author} {\bibinfo {author} {\bibfnamefont {M.}~\bibnamefont
  {Mannini}}, \bibinfo {author} {\bibfnamefont {F.}~\bibnamefont {Pineider}},
  \bibinfo {author} {\bibfnamefont {P.}~\bibnamefont {Sainctavit}}, \bibinfo
  {author} {\bibfnamefont {C.}~\bibnamefont {Danieli}}, \bibinfo {author}
  {\bibfnamefont {E.}~\bibnamefont {Otero}}, \bibinfo {author} {\bibfnamefont
  {C.}~\bibnamefont {Sciancalepore}}, \bibinfo {author} {\bibfnamefont {A.~M.}\
  \bibnamefont {Talarico}}, \bibinfo {author} {\bibfnamefont {M.-A.}\
  \bibnamefont {Arrio}}, \bibinfo {author} {\bibfnamefont {A.}~\bibnamefont
  {Cornia}}, \bibinfo {author} {\bibfnamefont {D.}~\bibnamefont {Gatteschi}}, \
  and\ \bibinfo {author} {\bibfnamefont {R.}~\bibnamefont {Sessoli}},\
  }\bibfield  {title} {\enquote {\bibinfo {title} {{Magnetic memory of a
  single-molecule quantum magnet wired to a gold surface}},}\ }\href {\doibase
  10.1038/nmat2374} {\bibfield  {journal} {\bibinfo  {journal} {Nat. Mater.}\
  }\textbf {\bibinfo {volume} {8}},\ \bibinfo {pages} {194} (\bibinfo {year}
  {2009})}\BibitemShut {NoStop}%
\bibitem [{\citenamefont {Vincent}\ \emph {et~al.}(2012)\citenamefont
  {Vincent}, \citenamefont {Klyatskaya}, \citenamefont {Ruben}, \citenamefont
  {Wernsdorfer},\ and\ \citenamefont {Balestro}}]{Vincent12}%
  \BibitemOpen
  \bibfield  {author} {\bibinfo {author} {\bibfnamefont {R.}~\bibnamefont
  {Vincent}}, \bibinfo {author} {\bibfnamefont {S.}~\bibnamefont {Klyatskaya}},
  \bibinfo {author} {\bibfnamefont {M.}~\bibnamefont {Ruben}}, \bibinfo
  {author} {\bibfnamefont {W.}~\bibnamefont {Wernsdorfer}}, \ and\ \bibinfo
  {author} {\bibfnamefont {F.}~\bibnamefont {Balestro}},\ }\bibfield  {title}
  {\enquote {\bibinfo {title} {{Electronic read-out of a single nuclear spin
  using a molecular spin transistor}},}\ }\href {\doibase 10.1038/nature11341}
  {\bibfield  {journal} {\bibinfo  {journal} {Nature}\ }\textbf {\bibinfo
  {volume} {488}},\ \bibinfo {pages} {357} (\bibinfo {year}
  {2012})}\BibitemShut {NoStop}%
\bibitem [{\citenamefont {Natterer}\ \emph {et~al.}(2017)\citenamefont
  {Natterer}, \citenamefont {Yang}, \citenamefont {Paul}, \citenamefont
  {Willke}, \citenamefont {Choi}, \citenamefont {Greber}, \citenamefont
  {Heinrich},\ and\ \citenamefont {Lutz}}]{Natterer17}%
  \BibitemOpen
  \bibfield  {author} {\bibinfo {author} {\bibfnamefont {F.~D.}\ \bibnamefont
  {Natterer}}, \bibinfo {author} {\bibfnamefont {K.}~\bibnamefont {Yang}},
  \bibinfo {author} {\bibfnamefont {W.}~\bibnamefont {Paul}}, \bibinfo {author}
  {\bibfnamefont {P.}~\bibnamefont {Willke}}, \bibinfo {author} {\bibfnamefont
  {T.}~\bibnamefont {Choi}}, \bibinfo {author} {\bibfnamefont {T.}~\bibnamefont
  {Greber}}, \bibinfo {author} {\bibfnamefont {A.~J.}\ \bibnamefont
  {Heinrich}}, \ and\ \bibinfo {author} {\bibfnamefont {C.~P.}\ \bibnamefont
  {Lutz}},\ }\bibfield  {title} {\enquote {\bibinfo {title} {{Reading and
  writing single-atom magnets}},}\ }\href {\doibase 10.1038/nature21371}
  {\bibfield  {journal} {\bibinfo  {journal} {Nature}\ }\textbf {\bibinfo
  {volume} {543}},\ \bibinfo {pages} {226} (\bibinfo {year}
  {2017})}\BibitemShut {NoStop}%
\bibitem [{\citenamefont {Maune}\ \emph {et~al.}(2012)\citenamefont {Maune},
  \citenamefont {Borselli}, \citenamefont {Huang}, \citenamefont {Ladd},
  \citenamefont {Deelman}, \citenamefont {Holabird}, \citenamefont {Kiselev},
  \citenamefont {Alvarado-Rodriguez}, \citenamefont {Ross}, \citenamefont
  {Schmitz}, \citenamefont {Sokolich}, \citenamefont {Watson}, \citenamefont
  {Gyure},\ and\ \citenamefont {Hunter}}]{Maune2012Jan}%
  \BibitemOpen
  \bibfield  {author} {\bibinfo {author} {\bibfnamefont {B.~M.}\ \bibnamefont
  {Maune}}, \bibinfo {author} {\bibfnamefont {M.~G.}\ \bibnamefont {Borselli}},
  \bibinfo {author} {\bibfnamefont {B.}~\bibnamefont {Huang}}, \bibinfo
  {author} {\bibfnamefont {T.~D.}\ \bibnamefont {Ladd}}, \bibinfo {author}
  {\bibfnamefont {P.~W.}\ \bibnamefont {Deelman}}, \bibinfo {author}
  {\bibfnamefont {K.~S.}\ \bibnamefont {Holabird}}, \bibinfo {author}
  {\bibfnamefont {A.~A.}\ \bibnamefont {Kiselev}}, \bibinfo {author}
  {\bibfnamefont {I.}~\bibnamefont {Alvarado-Rodriguez}}, \bibinfo {author}
  {\bibfnamefont {R.~S.}\ \bibnamefont {Ross}}, \bibinfo {author}
  {\bibfnamefont {A.~E.}\ \bibnamefont {Schmitz}}, \bibinfo {author}
  {\bibfnamefont {M.}~\bibnamefont {Sokolich}}, \bibinfo {author}
  {\bibfnamefont {C.~A.}\ \bibnamefont {Watson}}, \bibinfo {author}
  {\bibfnamefont {M.~F.}\ \bibnamefont {Gyure}}, \ and\ \bibinfo {author}
  {\bibfnamefont {A.~T.}\ \bibnamefont {Hunter}},\ }\bibfield  {title}
  {\enquote {\bibinfo {title} {{Coherent singlet-triplet oscillations in a
  silicon-based double quantum dot}},}\ }\href {\doibase 10.1038/nature10707}
  {\bibfield  {journal} {\bibinfo  {journal} {Nature}\ }\textbf {\bibinfo
  {volume} {481}},\ \bibinfo {pages} {344} (\bibinfo {year}
  {2012})}\BibitemShut {NoStop}%
\bibitem [{\citenamefont {Malinowski}\ \emph {et~al.}(2018)\citenamefont
  {Malinowski}, \citenamefont {Martins}, \citenamefont {Smith}, \citenamefont
  {Bartlett}, \citenamefont {Doherty}, \citenamefont {Nissen}, \citenamefont
  {Fallahi}, \citenamefont {Gardner}, \citenamefont {Manfra}, \citenamefont
  {Marcus},\ and\ \citenamefont {Kuemmeth}}]{Malinowski2018Mar}%
  \BibitemOpen
  \bibfield  {author} {\bibinfo {author} {\bibfnamefont {F.~K.}\ \bibnamefont
  {Malinowski}}, \bibinfo {author} {\bibfnamefont {F.}~\bibnamefont {Martins}},
  \bibinfo {author} {\bibfnamefont {T.~B.}\ \bibnamefont {Smith}}, \bibinfo
  {author} {\bibfnamefont {S.~D.}\ \bibnamefont {Bartlett}}, \bibinfo {author}
  {\bibfnamefont {A.~C.}\ \bibnamefont {Doherty}}, \bibinfo {author}
  {\bibfnamefont {P.~D.}\ \bibnamefont {Nissen}}, \bibinfo {author}
  {\bibfnamefont {S.}~\bibnamefont {Fallahi}}, \bibinfo {author} {\bibfnamefont
  {G.~C.}\ \bibnamefont {Gardner}}, \bibinfo {author} {\bibfnamefont {M.~J.}\
  \bibnamefont {Manfra}}, \bibinfo {author} {\bibfnamefont {C.~M.}\
  \bibnamefont {Marcus}}, \ and\ \bibinfo {author} {\bibfnamefont
  {F.}~\bibnamefont {Kuemmeth}},\ }\bibfield  {title} {\enquote {\bibinfo
  {title} {{Spin of a Multielectron Quantum Dot and Its Interaction with a
  Neighboring Electron}},}\ }\href {\doibase 10.1103/PhysRevX.8.011045}
  {\bibfield  {journal} {\bibinfo  {journal} {Phys. Rev. X}\ }\textbf {\bibinfo
  {volume} {8}},\ \bibinfo {pages} {011045} (\bibinfo {year}
  {2018})}\BibitemShut {NoStop}%
\bibitem [{\citenamefont
  {Bo{\ifmmode\check{c}\else\v{c}\fi}a}(1999)}]{Boca_book}%
  \BibitemOpen
  \bibfield  {author} {\bibinfo {author} {\bibfnamefont {R.}~\bibnamefont
  {Bo{\ifmmode\check{c}\else\v{c}\fi}a}},\ }\href
  {https://www.sciencedirect.com/bookseries/current-methods-in-inorganic-chemistry/vol/1}
  {\emph {\bibinfo {title} {{Theoretical Foundations of Molecular
  Magnetism}}}},\ \bibinfo {series} {{Current Methods in Inorganic Chemistry}},
  Vol.~\bibinfo {volume} {1}\ (\bibinfo  {publisher} {Elsevier Science},\
  \bibinfo {address} {Lausanne},\ \bibinfo {year} {1999})\BibitemShut {NoStop}%
\bibitem [{\citenamefont {Lehmann}\ \emph {et~al.}(2009)\citenamefont
  {Lehmann}, \citenamefont {Gaita-Ari{\ifmmode\tilde{n}\else\~{n}\fi}o},
  \citenamefont {Coronado},\ and\ \citenamefont {Loss}}]{Lehmann2009Mar}%
  \BibitemOpen
  \bibfield  {author} {\bibinfo {author} {\bibfnamefont {J.}~\bibnamefont
  {Lehmann}}, \bibinfo {author} {\bibfnamefont {A.}~\bibnamefont
  {Gaita-Ari{\ifmmode\tilde{n}\else\~{n}\fi}o}}, \bibinfo {author}
  {\bibfnamefont {E.}~\bibnamefont {Coronado}}, \ and\ \bibinfo {author}
  {\bibfnamefont {D.}~\bibnamefont {Loss}},\ }\bibfield  {title} {\enquote
  {\bibinfo {title} {{Quantum computing with molecular spin systems}},}\ }\href
  {\doibase 10.1039/B810634G} {\bibfield  {journal} {\bibinfo  {journal} {J.
  Mater. Chem.}\ }\textbf {\bibinfo {volume} {19}},\ \bibinfo {pages} {1672}
  (\bibinfo {year} {2009})}\BibitemShut {NoStop}%
\bibitem [{\citenamefont {Aromi}\ \emph {et~al.}(2012)\citenamefont {Aromi},
  \citenamefont {Aguila}, \citenamefont {Gamez}, \citenamefont {Luis},\ and\
  \citenamefont {Roubeau}}]{Aromi2012}%
  \BibitemOpen
  \bibfield  {author} {\bibinfo {author} {\bibfnamefont {G.}~\bibnamefont
  {Aromi}}, \bibinfo {author} {\bibfnamefont {D.}~\bibnamefont {Aguila}},
  \bibinfo {author} {\bibfnamefont {P.}~\bibnamefont {Gamez}}, \bibinfo
  {author} {\bibfnamefont {F.}~\bibnamefont {Luis}}, \ and\ \bibinfo {author}
  {\bibfnamefont {O.}~\bibnamefont {Roubeau}},\ }\bibfield  {title} {\enquote
  {\bibinfo {title} {{Design of magnetic coordination complexes for quantum
  computing}},}\ }\href {\doibase 10.1039/c1cs15115k} {\bibfield  {journal}
  {\bibinfo  {journal} {Chem. Soc. Rev.}\ }\textbf {\bibinfo {volume} {41}},\
  \bibinfo {pages} {537} (\bibinfo {year} {2012})}\BibitemShut {NoStop}%
\bibitem [{\citenamefont {Khajetoorians}\ and\ \citenamefont
  {Heinrich}(2016)}]{Khajetoorians2016Apr}%
  \BibitemOpen
  \bibfield  {author} {\bibinfo {author} {\bibfnamefont {A.~A.}\ \bibnamefont
  {Khajetoorians}}\ and\ \bibinfo {author} {\bibfnamefont {A.~J.}\ \bibnamefont
  {Heinrich}},\ }\bibfield  {title} {\enquote {\bibinfo {title} {{Toward
  single-atom memory}},}\ }\href {\doibase 10.1126/science.aaf2481} {\bibfield
  {journal} {\bibinfo  {journal} {Science}\ }\textbf {\bibinfo {volume}
  {352}},\ \bibinfo {pages} {296} (\bibinfo {year} {2016})}\BibitemShut
  {NoStop}%
\bibitem [{\citenamefont {Gatteschi}\ \emph {et~al.}(2006)\citenamefont
  {Gatteschi}, \citenamefont {Sessoli},\ and\ \citenamefont
  {Villain}}]{Gatteschi_book}%
  \BibitemOpen
  \bibfield  {author} {\bibinfo {author} {\bibfnamefont {D.}~\bibnamefont
  {Gatteschi}}, \bibinfo {author} {\bibfnamefont {R.}~\bibnamefont {Sessoli}},
  \ and\ \bibinfo {author} {\bibfnamefont {J.}~\bibnamefont {Villain}},\ }\href
  {\doibase 10.1093/acprof:oso/9780198567530.001.0001} {\emph {\bibinfo {title}
  {{Molecular nanomagnets}}}}\ (\bibinfo  {publisher} {Oxford University
  Press},\ \bibinfo {address} {New York},\ \bibinfo {year} {2006})\BibitemShut
  {NoStop}%
\bibitem [{\citenamefont {Martinek}\ \emph {et~al.}(2003)\citenamefont
  {Martinek}, \citenamefont {Utsumi}, \citenamefont {Imamura}, \citenamefont
  {Barna{\ifmmode\acute{s}\else\'{s}\fi}}, \citenamefont {Maekawa},
  \citenamefont {K{\ifmmode\ddot{o}\else\"{o}\fi}nig},\ and\ \citenamefont
  {Sch{\ifmmode\ddot{o}\else\"{o}\fi}n}}]{MartinekTK}%
  \BibitemOpen
  \bibfield  {author} {\bibinfo {author} {\bibfnamefont {J.}~\bibnamefont
  {Martinek}}, \bibinfo {author} {\bibfnamefont {Y.}~\bibnamefont {Utsumi}},
  \bibinfo {author} {\bibfnamefont {H.}~\bibnamefont {Imamura}}, \bibinfo
  {author} {\bibfnamefont {J.}~\bibnamefont
  {Barna{\ifmmode\acute{s}\else\'{s}\fi}}}, \bibinfo {author} {\bibfnamefont
  {S.}~\bibnamefont {Maekawa}}, \bibinfo {author} {\bibfnamefont
  {J.}~\bibnamefont {K{\ifmmode\ddot{o}\else\"{o}\fi}nig}}, \ and\ \bibinfo
  {author} {\bibfnamefont {G.}~\bibnamefont
  {Sch{\ifmmode\ddot{o}\else\"{o}\fi}n}},\ }\bibfield  {title} {\enquote
  {\bibinfo {title} {{Kondo Effect in Quantum Dots Coupled to Ferromagnetic
  Leads}},}\ }\href {\doibase 10.1103/PhysRevLett.91.127203} {\bibfield
  {journal} {\bibinfo  {journal} {Phys. Rev. Lett.}\ }\textbf {\bibinfo
  {volume} {91}},\ \bibinfo {pages} {127203} (\bibinfo {year}
  {2003})}\BibitemShut {NoStop}%
\bibitem [{\citenamefont {Martinek}\ \emph {et~al.}(2005)\citenamefont
  {Martinek}, \citenamefont {Sindel}, \citenamefont {Borda}, \citenamefont
  {Barna{\ifmmode\acute{s}\else\'{s}\fi}}, \citenamefont {Bulla}, \citenamefont
  {K{\ifmmode\ddot{o}\else\"{o}\fi}nig}, \citenamefont
  {Sch{\ifmmode\ddot{o}\else\"{o}\fi}n}, \citenamefont {Maekawa},\ and\
  \citenamefont {von Delft}}]{MartinekEx}%
  \BibitemOpen
  \bibfield  {author} {\bibinfo {author} {\bibfnamefont {J.}~\bibnamefont
  {Martinek}}, \bibinfo {author} {\bibfnamefont {M.}~\bibnamefont {Sindel}},
  \bibinfo {author} {\bibfnamefont {L.}~\bibnamefont {Borda}}, \bibinfo
  {author} {\bibfnamefont {J.}~\bibnamefont
  {Barna{\ifmmode\acute{s}\else\'{s}\fi}}}, \bibinfo {author} {\bibfnamefont
  {R.}~\bibnamefont {Bulla}}, \bibinfo {author} {\bibfnamefont
  {J.}~\bibnamefont {K{\ifmmode\ddot{o}\else\"{o}\fi}nig}}, \bibinfo {author}
  {\bibfnamefont {G.}~\bibnamefont {Sch{\ifmmode\ddot{o}\else\"{o}\fi}n}},
  \bibinfo {author} {\bibfnamefont {S.}~\bibnamefont {Maekawa}}, \ and\
  \bibinfo {author} {\bibfnamefont {J.}~\bibnamefont {von Delft}},\ }\bibfield
  {title} {\enquote {\bibinfo {title} {{Gate-controlled spin splitting in
  quantum dots with ferromagnetic leads in the Kondo regime}},}\ }\href
  {\doibase 10.1103/PhysRevB.72.121302} {\bibfield  {journal} {\bibinfo
  {journal} {Phys. Rev. B}\ }\textbf {\bibinfo {volume} {72}},\ \bibinfo
  {pages} {121302(R)} (\bibinfo {year} {2005})}\BibitemShut {NoStop}%
\bibitem [{\citenamefont {Hauptmann}\ \emph {et~al.}(2008)\citenamefont
  {Hauptmann}, \citenamefont {Paaske},\ and\ \citenamefont
  {Lindelof}}]{Hauptmann_Nat.Phys.4/2008}%
  \BibitemOpen
  \bibfield  {author} {\bibinfo {author} {\bibfnamefont {J.~R.}\ \bibnamefont
  {Hauptmann}}, \bibinfo {author} {\bibfnamefont {J.}~\bibnamefont {Paaske}}, \
  and\ \bibinfo {author} {\bibfnamefont {P.~E.}\ \bibnamefont {Lindelof}},\
  }\bibfield  {title} {\enquote {\bibinfo {title} {{Electric-field-controlled
  spin reversal in a quantum dot with ferromagnetic contacts}},}\ }\href
  {\doibase 10.1038/nphys931} {\bibfield  {journal} {\bibinfo  {journal} {Nat.
  Phys.}\ }\textbf {\bibinfo {volume} {4}},\ \bibinfo {pages} {373} (\bibinfo
  {year} {2008})}\BibitemShut {NoStop}%
\bibitem [{\citenamefont {Gaass}\ \emph {et~al.}(2011)\citenamefont {Gaass},
  \citenamefont {H{\ifmmode\ddot{u}\else\"{u}\fi}ttel}, \citenamefont {Kang},
  \citenamefont {Weymann}, \citenamefont {von Delft},\ and\ \citenamefont
  {Strunk}}]{Gaass_Phys.Rev.Lett.107/2011}%
  \BibitemOpen
  \bibfield  {author} {\bibinfo {author} {\bibfnamefont {M.}~\bibnamefont
  {Gaass}}, \bibinfo {author} {\bibfnamefont {A.~K.}\ \bibnamefont
  {H{\ifmmode\ddot{u}\else\"{u}\fi}ttel}}, \bibinfo {author} {\bibfnamefont
  {K.}~\bibnamefont {Kang}}, \bibinfo {author} {\bibfnamefont {I.}~\bibnamefont
  {Weymann}}, \bibinfo {author} {\bibfnamefont {J.}~\bibnamefont {von Delft}},
  \ and\ \bibinfo {author} {\bibfnamefont {{\relax Ch}.}~\bibnamefont
  {Strunk}},\ }\bibfield  {title} {\enquote {\bibinfo {title} {{Universality of
  the Kondo Effect in Quantum Dots with Ferromagnetic Leads}},}\ }\href
  {\doibase 10.1103/PhysRevLett.107.176808} {\bibfield  {journal} {\bibinfo
  {journal} {Phys. Rev. Lett.}\ }\textbf {\bibinfo {volume} {107}},\ \bibinfo
  {pages} {176808} (\bibinfo {year} {2011})}\BibitemShut {NoStop}%
\bibitem [{\citenamefont {Misiorny}\ \emph {et~al.}(2012)\citenamefont
  {Misiorny}, \citenamefont {Weymann},\ and\ \citenamefont
  {Barna{\ifmmode\acute{s}\else\'{s}\fi}}}]{Misiorny2012Dec}%
  \BibitemOpen
  \bibfield  {author} {\bibinfo {author} {\bibfnamefont {M.}~\bibnamefont
  {Misiorny}}, \bibinfo {author} {\bibfnamefont {I.}~\bibnamefont {Weymann}}, \
  and\ \bibinfo {author} {\bibfnamefont {J.}~\bibnamefont
  {Barna{\ifmmode\acute{s}\else\'{s}\fi}}},\ }\bibfield  {title} {\enquote
  {\bibinfo {title} {{Underscreened Kondo effect in $S=1$ magnetic quantum
  dots: Exchange, anisotropy, and temperature effects}},}\ }\href {\doibase
  10.1103/PhysRevB.86.245415} {\bibfield  {journal} {\bibinfo  {journal} {Phys.
  Rev. B}\ }\textbf {\bibinfo {volume} {86}},\ \bibinfo {pages} {245415}
  (\bibinfo {year} {2012})}\BibitemShut {NoStop}%
\bibitem [{\citenamefont {Misiorny}\ \emph {et~al.}(2013)\citenamefont
  {Misiorny}, \citenamefont {Hell},\ and\ \citenamefont
  {Wegewijs}}]{Misiorny2013}%
  \BibitemOpen
  \bibfield  {author} {\bibinfo {author} {\bibfnamefont {M.}~\bibnamefont
  {Misiorny}}, \bibinfo {author} {\bibfnamefont {M.}~\bibnamefont {Hell}}, \
  and\ \bibinfo {author} {\bibfnamefont {M.~R.}\ \bibnamefont {Wegewijs}},\
  }\bibfield  {title} {\enquote {\bibinfo {title} {{Spintronic magnetic
  anisotropy}},}\ }\href {\doibase 10.1038/nphys2766} {\bibfield  {journal}
  {\bibinfo  {journal} {Nat. Phys.}\ }\textbf {\bibinfo {volume} {9}},\
  \bibinfo {pages} {801} (\bibinfo {year} {2013})}\BibitemShut {NoStop}%
\bibitem [{\citenamefont {Hatter}\ \emph {et~al.}(2015)\citenamefont {Hatter},
  \citenamefont {Heinrich}, \citenamefont {Ruby}, \citenamefont {Pascual},\
  and\ \citenamefont {Franke}}]{YSRanisotropy}%
  \BibitemOpen
  \bibfield  {author} {\bibinfo {author} {\bibfnamefont {N.}~\bibnamefont
  {Hatter}}, \bibinfo {author} {\bibfnamefont {B.~W.}\ \bibnamefont
  {Heinrich}}, \bibinfo {author} {\bibfnamefont {M.}~\bibnamefont {Ruby}},
  \bibinfo {author} {\bibfnamefont {J.~I.}\ \bibnamefont {Pascual}}, \ and\
  \bibinfo {author} {\bibfnamefont {K.~J.}\ \bibnamefont {Franke}},\ }\bibfield
   {title} {\enquote {\bibinfo {title} {{Magnetic anisotropy in Shiba bound
  states across a quantum phase transition}},}\ }\href {\doibase
  10.1038/ncomms9988} {\bibfield  {journal} {\bibinfo  {journal} {Nat.
  Commun.}\ }\textbf {\bibinfo {volume} {6}},\ \bibinfo {pages} {8988}
  (\bibinfo {year} {2015})}\BibitemShut {NoStop}%
\bibitem [{\citenamefont {Sessoli}(2017)}]{Sessoli17}%
  \BibitemOpen
  \bibfield  {author} {\bibinfo {author} {\bibfnamefont {R.}~\bibnamefont
  {Sessoli}},\ }\bibfield  {title} {\enquote {\bibinfo {title} {{Magnetic
  molecules back in the race}},}\ }\href {\doibase 10.1038/548400a} {\bibfield
  {journal} {\bibinfo  {journal} {Nature}\ }\textbf {\bibinfo {volume} {548}},\
  \bibinfo {pages} {400} (\bibinfo {year} {2017})}\BibitemShut {NoStop}%
\bibitem [{\citenamefont {De~Franceschi}\ \emph {et~al.}(2010)\citenamefont
  {De~Franceschi}, \citenamefont {Kouwenhoven}, \citenamefont
  {Sch{\ifmmode\ddot{o}\else\"{o}\fi}nenberger},\ and\ \citenamefont
  {Wernsdorfer}}]{hybridQDs}%
  \BibitemOpen
  \bibfield  {author} {\bibinfo {author} {\bibfnamefont {S.}~\bibnamefont
  {De~Franceschi}}, \bibinfo {author} {\bibfnamefont {L.}~\bibnamefont
  {Kouwenhoven}}, \bibinfo {author} {\bibfnamefont {C.}~\bibnamefont
  {Sch{\ifmmode\ddot{o}\else\"{o}\fi}nenberger}}, \ and\ \bibinfo {author}
  {\bibfnamefont {W.}~\bibnamefont {Wernsdorfer}},\ }\bibfield  {title}
  {\enquote {\bibinfo {title} {{Hybrid superconductor{\textendash}quantum dot
  devices}},}\ }\href {\doibase 10.1038/nnano.2010.173} {\bibfield  {journal}
  {\bibinfo  {journal} {Nat. Nanotechnol.}\ }\textbf {\bibinfo {volume} {5}},\
  \bibinfo {pages} {703} (\bibinfo {year} {2010})}\BibitemShut {NoStop}%
\bibitem [{\citenamefont {Martin-Rodero}\ and\ \citenamefont
  {Yeyati}(2011)}]{MartinRodero_AdinPhys2011}%
  \BibitemOpen
  \bibfield  {author} {\bibinfo {author} {\bibfnamefont {A.}~\bibnamefont
  {Martin-Rodero}}\ and\ \bibinfo {author} {\bibfnamefont {A.~Levy}\
  \bibnamefont {Yeyati}},\ }\bibfield  {title} {\enquote {\bibinfo {title}
  {Josephson and andreev transport through quantum dots},}\ }\href {\doibase
  10.1080/00018732.2011.624266} {\bibfield  {journal} {\bibinfo  {journal}
  {Advances in Physics}\ }\textbf {\bibinfo {volume} {60}},\ \bibinfo {pages}
  {899} (\bibinfo {year} {2011})}\BibitemShut {NoStop}%
\bibitem [{\citenamefont {Yu}(1965)}]{Yu}%
  \BibitemOpen
  \bibfield  {author} {\bibinfo {author} {\bibfnamefont {L.}~\bibnamefont
  {Yu}},\ }\bibfield  {title} {\enquote {\bibinfo {title} {{Bound state in
  superconductors with paramagnetic impurities}},}\ }\href
  {http://en.cnki.com.cn/Article_en/CJFDTOTAL-WLXB196501007.htm} {\bibfield
  {journal} {\bibinfo  {journal} {Acta Phys. Sin.}\ }\textbf {\bibinfo {volume}
  {21}},\ \bibinfo {pages} {75} (\bibinfo {year} {1965})}\BibitemShut {NoStop}%
\bibitem [{\citenamefont {Shiba}(1968)}]{Shiba}%
  \BibitemOpen
  \bibfield  {author} {\bibinfo {author} {\bibfnamefont {H.}~\bibnamefont
  {Shiba}},\ }\bibfield  {title} {\enquote {\bibinfo {title} {{Classical spins
  in superconductors}},}\ }\href
  {https://academic.oup.com/ptp/article/40/3/435/1831894} {\bibfield  {journal}
  {\bibinfo  {journal} {Prog. Teor. Phys.}\ }\textbf {\bibinfo {volume} {40}},\
  \bibinfo {pages} {435} (\bibinfo {year} {1968})}\BibitemShut {NoStop}%
\bibitem [{\citenamefont {Rusinov}(1969)}]{Rusinov}%
  \BibitemOpen
  \bibfield  {author} {\bibinfo {author} {\bibfnamefont {A.~I.}\ \bibnamefont
  {Rusinov}},\ }\bibfield  {title} {\enquote {\bibinfo {title}
  {{Superconductivity near a paramagnetic impurity}},}\ }\href
  {http://www.jetpletters.ac.ru/ps/1658/article_25295.shtml} {\bibfield
  {journal} {\bibinfo  {journal} {J. Exp. Teor. Phys. Lett.}\ }\textbf
  {\bibinfo {volume} {9}},\ \bibinfo {pages} {85} (\bibinfo {year} {1969})},\
  \bibinfo {note} {[Pisma Zh. Eksp. Teor. Fiz. 9, 146 (1968)]}\BibitemShut
  {NoStop}%
\bibitem [{\citenamefont {Lee}\ \emph {et~al.}(2014)\citenamefont {Lee},
  \citenamefont {Jiang}, \citenamefont {Houzet}, \citenamefont {Aguado},
  \citenamefont {Lieber},\ and\ \citenamefont {De~Franceschi}}]{Lee2014Jan}%
  \BibitemOpen
  \bibfield  {author} {\bibinfo {author} {\bibfnamefont {E.~J.~H.}\
  \bibnamefont {Lee}}, \bibinfo {author} {\bibfnamefont {X.}~\bibnamefont
  {Jiang}}, \bibinfo {author} {\bibfnamefont {M.}~\bibnamefont {Houzet}},
  \bibinfo {author} {\bibfnamefont {R.}~\bibnamefont {Aguado}}, \bibinfo
  {author} {\bibfnamefont {C.~M.}\ \bibnamefont {Lieber}}, \ and\ \bibinfo
  {author} {\bibfnamefont {S.}~\bibnamefont {De~Franceschi}},\ }\bibfield
  {title} {\enquote {\bibinfo {title} {Spin-resolved andreev levels and parity
  crossings in hybrid superconductor-semiconductor nanostructures},}\ }\href
  {\doibase 10.1038/nnano.2013.267} {\bibfield  {journal} {\bibinfo  {journal}
  {Nat. Nanotechnol.}\ }\textbf {\bibinfo {volume} {9}},\ \bibinfo {pages} {79}
  (\bibinfo {year} {2014})}\BibitemShut {NoStop}%
\bibitem [{\citenamefont {Island}\ \emph {et~al.}(2017)\citenamefont {Island},
  \citenamefont {Gaudenzi}, \citenamefont {de~Bruijckere}, \citenamefont
  {Burzur{\ifmmode\acute{\imath}\else\'{\i}\fi}}, \citenamefont {Franco},
  \citenamefont {Mas-Torrent}, \citenamefont {Rovira}, \citenamefont {Veciana},
  \citenamefont {Klapwijk}, \citenamefont {Aguado},\ and\ \citenamefont
  {van~der Zant}}]{vdZant_inducedShiba}%
  \BibitemOpen
  \bibfield  {author} {\bibinfo {author} {\bibfnamefont {J.~O.}\ \bibnamefont
  {Island}}, \bibinfo {author} {\bibfnamefont {R.}~\bibnamefont {Gaudenzi}},
  \bibinfo {author} {\bibfnamefont {J.}~\bibnamefont {de~Bruijckere}}, \bibinfo
  {author} {\bibfnamefont {E.}~\bibnamefont
  {Burzur{\ifmmode\acute{\imath}\else\'{\i}\fi}}}, \bibinfo {author}
  {\bibfnamefont {C.}~\bibnamefont {Franco}}, \bibinfo {author} {\bibfnamefont
  {M.}~\bibnamefont {Mas-Torrent}}, \bibinfo {author} {\bibfnamefont
  {C.}~\bibnamefont {Rovira}}, \bibinfo {author} {\bibfnamefont
  {J.}~\bibnamefont {Veciana}}, \bibinfo {author} {\bibfnamefont {T.~M.}\
  \bibnamefont {Klapwijk}}, \bibinfo {author} {\bibfnamefont {R.}~\bibnamefont
  {Aguado}}, \ and\ \bibinfo {author} {\bibfnamefont {H.~S.~J.}\ \bibnamefont
  {van~der Zant}},\ }\bibfield  {title} {\enquote {\bibinfo {title}
  {{Proximity-Induced Shiba States in a Molecular Junction}},}\ }\href
  {\doibase 10.1103/PhysRevLett.118.117001} {\bibfield  {journal} {\bibinfo
  {journal} {Phys. Rev. Lett.}\ }\textbf {\bibinfo {volume} {118}},\ \bibinfo
  {pages} {117001} (\bibinfo {year} {2017})}\BibitemShut {NoStop}%
\bibitem [{\citenamefont {Grove-Rasmussen}\ \emph {et~al.}(2018)\citenamefont
  {Grove-Rasmussen}, \citenamefont {Steffensen}, \citenamefont {Jellinggaard},
  \citenamefont {Madsen}, \citenamefont {{\ifmmode\check{Z}\else\v{Z}\fi}itko},
  \citenamefont {Paaske},\ and\ \citenamefont {Nyg{\aa}rd}}]{YSRscreening}%
  \BibitemOpen
  \bibfield  {author} {\bibinfo {author} {\bibfnamefont {K.}~\bibnamefont
  {Grove-Rasmussen}}, \bibinfo {author} {\bibfnamefont {G.}~\bibnamefont
  {Steffensen}}, \bibinfo {author} {\bibfnamefont {A.}~\bibnamefont
  {Jellinggaard}}, \bibinfo {author} {\bibfnamefont {M.~H.}\ \bibnamefont
  {Madsen}}, \bibinfo {author} {\bibfnamefont {R.}~\bibnamefont
  {{\ifmmode\check{Z}\else\v{Z}\fi}itko}}, \bibinfo {author} {\bibfnamefont
  {J.}~\bibnamefont {Paaske}}, \ and\ \bibinfo {author} {\bibfnamefont
  {J.}~\bibnamefont {Nyg{\aa}rd}},\ }\bibfield  {title} {\enquote {\bibinfo
  {title} {{Yu{\textendash}Shiba{\textendash}Rusinov screening of spins in
  double quantum dots}},}\ }\href {\doibase 10.1038/s41467-018-04683-x}
  {\bibfield  {journal} {\bibinfo  {journal} {Nat. Commun.}\ }\textbf {\bibinfo
  {volume} {9}},\ \bibinfo {pages} {2376} (\bibinfo {year} {2018})}\BibitemShut
  {NoStop}%
\bibitem [{\citenamefont {Hofstetter}\ \emph {et~al.}(2009)\citenamefont
  {Hofstetter}, \citenamefont {Csonka}, \citenamefont {Nyg{\aa}rd},\ and\
  \citenamefont {Sch{\ifmmode\ddot{o}\else\"{o}\fi}nenberger}}]{CPS1}%
  \BibitemOpen
  \bibfield  {author} {\bibinfo {author} {\bibfnamefont {L.}~\bibnamefont
  {Hofstetter}}, \bibinfo {author} {\bibfnamefont {S.}~\bibnamefont {Csonka}},
  \bibinfo {author} {\bibfnamefont {J.}~\bibnamefont {Nyg{\aa}rd}}, \ and\
  \bibinfo {author} {\bibfnamefont {C.}~\bibnamefont
  {Sch{\ifmmode\ddot{o}\else\"{o}\fi}nenberger}},\ }\bibfield  {title}
  {\enquote {\bibinfo {title} {{Cooper pair splitter realized in a
  two-quantum-dot Y-junction}},}\ }\href {\doibase 10.1038/nature08432}
  {\bibfield  {journal} {\bibinfo  {journal} {Nature}\ }\textbf {\bibinfo
  {volume} {461}},\ \bibinfo {pages} {960} (\bibinfo {year}
  {2009})}\BibitemShut {NoStop}%
\bibitem [{\citenamefont {Borzenets}\ \emph {et~al.}(2016)\citenamefont
  {Borzenets}, \citenamefont {Shimazaki}, \citenamefont {Jones}, \citenamefont
  {Craciun}, \citenamefont {Russo}, \citenamefont {Yamamoto},\ and\
  \citenamefont {Tarucha}}]{CPS9}%
  \BibitemOpen
  \bibfield  {author} {\bibinfo {author} {\bibfnamefont {I.~V.}\ \bibnamefont
  {Borzenets}}, \bibinfo {author} {\bibfnamefont {Y.}~\bibnamefont
  {Shimazaki}}, \bibinfo {author} {\bibfnamefont {G.~F.}\ \bibnamefont
  {Jones}}, \bibinfo {author} {\bibfnamefont {M.~F.}\ \bibnamefont {Craciun}},
  \bibinfo {author} {\bibfnamefont {S.}~\bibnamefont {Russo}}, \bibinfo
  {author} {\bibfnamefont {M.}~\bibnamefont {Yamamoto}}, \ and\ \bibinfo
  {author} {\bibfnamefont {S.}~\bibnamefont {Tarucha}},\ }\bibfield  {title}
  {\enquote {\bibinfo {title} {{High Efficiency CVD Graphene-lead (Pb) Cooper
  Pair Splitter}},}\ }\href {\doibase 10.1038/srep23051} {\bibfield  {journal}
  {\bibinfo  {journal} {Sci. Rep.}\ }\textbf {\bibinfo {volume} {6}},\ \bibinfo
  {pages} {23051} (\bibinfo {year} {2016})}\BibitemShut {NoStop}%
\bibitem [{\citenamefont {Franke}\ \emph {et~al.}(2011)\citenamefont {Franke},
  \citenamefont {Schulze},\ and\ \citenamefont {Pascual}}]{Franke_Science11}%
  \BibitemOpen
  \bibfield  {author} {\bibinfo {author} {\bibfnamefont {K.~J.}\ \bibnamefont
  {Franke}}, \bibinfo {author} {\bibfnamefont {G.}~\bibnamefont {Schulze}}, \
  and\ \bibinfo {author} {\bibfnamefont {J.~I.}\ \bibnamefont {Pascual}},\
  }\bibfield  {title} {\enquote {\bibinfo {title} {{Competition of
  Superconducting Phenomena and Kondo Screening at the Nanoscale}},}\ }\href
  {\doibase 10.1126/science.1202204} {\bibfield  {journal} {\bibinfo  {journal}
  {Science}\ }\textbf {\bibinfo {volume} {332}},\ \bibinfo {pages} {940}
  (\bibinfo {year} {2011})}\BibitemShut {NoStop}%
\bibitem [{\citenamefont {Pillet}\ \emph {et~al.}(2013)\citenamefont {Pillet},
  \citenamefont {Joyez}, \citenamefont {{\ifmmode \check{Z} \else
  \v{Z}\fi}itko},\ and\ \citenamefont {Goffman}}]{Pillet2013Jul}%
  \BibitemOpen
  \bibfield  {author} {\bibinfo {author} {\bibfnamefont {J.-D.}\ \bibnamefont
  {Pillet}}, \bibinfo {author} {\bibfnamefont {P.}~\bibnamefont {Joyez}},
  \bibinfo {author} {\bibfnamefont {Rok}\ \bibnamefont {{\ifmmode \check{Z}
  \else \v{Z}\fi}itko}}, \ and\ \bibinfo {author} {\bibfnamefont {M.~F.}\
  \bibnamefont {Goffman}},\ }\bibfield  {title} {\enquote {\bibinfo {title}
  {Tunneling spectroscopy of a single quantum dot coupled to a superconductor:
  From kondo ridge to andreev bound states},}\ }\href {\doibase
  10.1103/PhysRevB.88.045101} {\bibfield  {journal} {\bibinfo  {journal} {Phys.
  Rev. B}\ }\textbf {\bibinfo {volume} {88}},\ \bibinfo {pages} {045101}
  (\bibinfo {year} {2013})}\BibitemShut {NoStop}%
\bibitem [{\citenamefont {Doma{\ifmmode\acute{n}\else\'{n}\fi}ski}\ \emph
  {et~al.}(2016)\citenamefont {Doma{\ifmmode\acute{n}\else\'{n}\fi}ski},
  \citenamefont {Weymann}, \citenamefont
  {Bara{\ifmmode\acute{n}\else\'{n}\fi}ska},\ and\ \citenamefont
  {G{\ifmmode\acute{o}\else\'{o}\fi}rski}}]{Domanski-IW}%
  \BibitemOpen
  \bibfield  {author} {\bibinfo {author} {\bibfnamefont {T.}~\bibnamefont
  {Doma{\ifmmode\acute{n}\else\'{n}\fi}ski}}, \bibinfo {author} {\bibfnamefont
  {I.}~\bibnamefont {Weymann}}, \bibinfo {author} {\bibfnamefont
  {M.}~\bibnamefont {Bara{\ifmmode\acute{n}\else\'{n}\fi}ska}}, \ and\ \bibinfo
  {author} {\bibfnamefont {G.}~\bibnamefont
  {G{\ifmmode\acute{o}\else\'{o}\fi}rski}},\ }\bibfield  {title} {\enquote
  {\bibinfo {title} {{Constructive influence of the induced electron pairing on
  the Kondo state}},}\ }\href {\doibase 10.1038/srep23336} {\bibfield
  {journal} {\bibinfo  {journal} {Sci. Rep.}\ }\textbf {\bibinfo {volume}
  {6}},\ \bibinfo {pages} {23336} (\bibinfo {year} {2016})}\BibitemShut
  {NoStop}%
\bibitem [{\citenamefont {Wilson}(1975)}]{WilsonNRG}%
  \BibitemOpen
  \bibfield  {author} {\bibinfo {author} {\bibfnamefont {K.~G.}\ \bibnamefont
  {Wilson}},\ }\bibfield  {title} {\enquote {\bibinfo {title} {{The
  renormalization group: Critical phenomena and the Kondo problem}},}\ }\href
  {\doibase 10.1103/RevModPhys.47.773} {\bibfield  {journal} {\bibinfo
  {journal} {Rev. Mod. Phys.}\ }\textbf {\bibinfo {volume} {47}},\ \bibinfo
  {pages} {773} (\bibinfo {year} {1975})}\BibitemShut {NoStop}%
\bibitem [{\citenamefont {Misiorny}\ \emph {et~al.}(2015)\citenamefont
  {Misiorny}, \citenamefont {Burzur{\ifmmode\acute{\imath}\else\'{\i}\fi}},
  \citenamefont {Gaudenzi}, \citenamefont {Park}, \citenamefont {Leijnse},
  \citenamefont {Wegewijs}, \citenamefont {Paaske}, \citenamefont {Cornia},\
  and\ \citenamefont {van~der Zant}}]{Misiorny_vdZant}%
  \BibitemOpen
  \bibfield  {author} {\bibinfo {author} {\bibfnamefont {M.}~\bibnamefont
  {Misiorny}}, \bibinfo {author} {\bibfnamefont {E.}~\bibnamefont
  {Burzur{\ifmmode\acute{\imath}\else\'{\i}\fi}}}, \bibinfo {author}
  {\bibfnamefont {R.}~\bibnamefont {Gaudenzi}}, \bibinfo {author}
  {\bibfnamefont {K.}~\bibnamefont {Park}}, \bibinfo {author} {\bibfnamefont
  {M.}~\bibnamefont {Leijnse}}, \bibinfo {author} {\bibfnamefont {M.~R.}\
  \bibnamefont {Wegewijs}}, \bibinfo {author} {\bibfnamefont {J.}~\bibnamefont
  {Paaske}}, \bibinfo {author} {\bibfnamefont {A.}~\bibnamefont {Cornia}}, \
  and\ \bibinfo {author} {\bibfnamefont {H.~S.~J.}\ \bibnamefont {van~der
  Zant}},\ }\bibfield  {title} {\enquote {\bibinfo {title} {{Probing transverse
  magnetic anisotropy by electronic transport through a single-molecule
  magnet}},}\ }\href {\doibase 10.1103/PhysRevB.91.035442} {\bibfield
  {journal} {\bibinfo  {journal} {Phys. Rev. B}\ }\textbf {\bibinfo {volume}
  {91}},\ \bibinfo {pages} {035442} (\bibinfo {year} {2015})}\BibitemShut
  {NoStop}%
\bibitem [{\citenamefont {Rozhkov}\ and\ \citenamefont
  {Arovas}(2000)}]{RozhkovArovas}%
  \BibitemOpen
  \bibfield  {author} {\bibinfo {author} {\bibfnamefont {A.~V.}\ \bibnamefont
  {Rozhkov}}\ and\ \bibinfo {author} {\bibfnamefont {D.~P.}\ \bibnamefont
  {Arovas}},\ }\bibfield  {title} {\enquote {\bibinfo {title}
  {{Interacting-impurity Josephson junction: Variational wave functions and
  slave-boson mean-field theory}},}\ }\href {\doibase 10.1103/PhysRevB.62.6687}
  {\bibfield  {journal} {\bibinfo  {journal} {Phys. Rev. B}\ }\textbf {\bibinfo
  {volume} {62}},\ \bibinfo {pages} {6687} (\bibinfo {year}
  {2000})}\BibitemShut {NoStop}%
\bibitem [{\citenamefont {Buitelaar}\ \emph {et~al.}(2003)\citenamefont
  {Buitelaar}, \citenamefont {Belzig}, \citenamefont {Nussbaumer},
  \citenamefont {Babi{\ifmmode\acute{c}\else\'{c}\fi}}, \citenamefont
  {Bruder},\ and\ \citenamefont
  {Sch{\ifmmode\ddot{o}\else\"{o}\fi}nenberger}}]{Buitelaar}%
  \BibitemOpen
  \bibfield  {author} {\bibinfo {author} {\bibfnamefont {M.~R.}\ \bibnamefont
  {Buitelaar}}, \bibinfo {author} {\bibfnamefont {W.}~\bibnamefont {Belzig}},
  \bibinfo {author} {\bibfnamefont {T.}~\bibnamefont {Nussbaumer}}, \bibinfo
  {author} {\bibfnamefont {B.}~\bibnamefont
  {Babi{\ifmmode\acute{c}\else\'{c}\fi}}}, \bibinfo {author} {\bibfnamefont
  {C.}~\bibnamefont {Bruder}}, \ and\ \bibinfo {author} {\bibfnamefont
  {C.}~\bibnamefont {Sch{\ifmmode\ddot{o}\else\"{o}\fi}nenberger}},\ }\bibfield
   {title} {\enquote {\bibinfo {title} {{Multiple Andreev Reflections in a
  Carbon Nanotube Quantum Dot}},}\ }\href {\doibase
  10.1103/PhysRevLett.91.057005} {\bibfield  {journal} {\bibinfo  {journal}
  {Phys. Rev. Lett.}\ }\textbf {\bibinfo {volume} {91}},\ \bibinfo {pages}
  {057005} (\bibinfo {year} {2003})}\BibitemShut {NoStop}%
\bibitem [{\citenamefont {Pawlicki}\ and\ \citenamefont
  {Weymann}(2018)}]{Pawlicki2018Aug}%
  \BibitemOpen
  \bibfield  {author} {\bibinfo {author} {\bibfnamefont {F.}~\bibnamefont
  {Pawlicki}}\ and\ \bibinfo {author} {\bibfnamefont {I.}~\bibnamefont
  {Weymann}},\ }\bibfield  {title} {\enquote {\bibinfo {title} {{Andreev
  transport through single-molecule magnets}},}\ }\href {\doibase
  10.1103/PhysRevB.98.085411} {\bibfield  {journal} {\bibinfo  {journal} {Phys.
  Rev. B}\ }\textbf {\bibinfo {volume} {98}},\ \bibinfo {pages} {085411}
  (\bibinfo {year} {2018})}\BibitemShut {NoStop}%
\bibitem [{\citenamefont {Glazman}\ and\ \citenamefont
  {Raikh}(1988)}]{even-odd}%
  \BibitemOpen
  \bibfield  {author} {\bibinfo {author} {\bibfnamefont {L.~I.}\ \bibnamefont
  {Glazman}}\ and\ \bibinfo {author} {\bibfnamefont {M.~E.}\ \bibnamefont
  {Raikh}},\ }\bibfield  {title} {\enquote {\bibinfo {title} {{Resonant Kondo
  transparency of a barrier with quasilocal impurity states}},}\ }\href
  {http://www.jetpletters.ac.ru/ps/1095/article_16538.shtml} {\bibfield
  {journal} {\bibinfo  {journal} {J. Exp. Theor. Phys. Lett.}\ }\textbf
  {\bibinfo {volume} {47}},\ \bibinfo {pages} {452} (\bibinfo {year}
  {1988})}\BibitemShut {NoStop}%
\bibitem [{\citenamefont {Legeza}\ \emph {et~al.}(2008)\citenamefont {Legeza},
  \citenamefont {Moca}, \citenamefont {T\'{o}th}, \citenamefont {Weymann},\
  and\ \citenamefont {Zar\'{a}nd}}]{fnrg}%
  \BibitemOpen
  \bibfield  {author} {\bibinfo {author} {\bibfnamefont {\"{O}.}\ \bibnamefont
  {Legeza}}, \bibinfo {author} {\bibfnamefont {C.~P.}\ \bibnamefont {Moca}},
  \bibinfo {author} {\bibfnamefont {A.~I.}\ \bibnamefont {T\'{o}th}}, \bibinfo
  {author} {\bibfnamefont {I.}~\bibnamefont {Weymann}}, \ and\ \bibinfo
  {author} {\bibfnamefont {G.}~\bibnamefont {Zar\'{a}nd}},\ }\href
  {http://arxiv.org/abs/0809.3143} {\enquote {\bibinfo {title} {{Manual for the
  Flexible DM-NRG code}},}\ }\bibinfo {howpublished} {arXiv:0809.3143v1}
  (\bibinfo {year} {2008}),\ \bibinfo {note} {(the open access Flexible DM-NRG
  Budapest code is available at
  \href{http://www.phy.bme.hu/\~dmnrg/}{http:/\!/www.phy.bme.hu/\textasciitilde{}dmnrg/}}\BibitemShut
  {NoStop}%
\bibitem [{\citenamefont {Oliveira}\ and\ \citenamefont {Oliveira}(1994)}]{Z}%
  \BibitemOpen
  \bibfield  {author} {\bibinfo {author} {\bibfnamefont {W.~C.}\ \bibnamefont
  {Oliveira}}\ and\ \bibinfo {author} {\bibfnamefont {L.~N.}\ \bibnamefont
  {Oliveira}},\ }\bibfield  {title} {\enquote {\bibinfo {title} {{Generalized
  numerical renormalization-group method to calculate the thermodynamical
  properties of impurities in metals}},}\ }\href {\doibase
  10.1103/PhysRevB.49.11986} {\bibfield  {journal} {\bibinfo  {journal} {Phys.
  Rev. B}\ }\textbf {\bibinfo {volume} {49}},\ \bibinfo {pages} {11986}
  (\bibinfo {year} {1994})}\BibitemShut {NoStop}%
\bibitem [{\citenamefont {Hewson}(1997)}]{Hewson_book}%
  \BibitemOpen
  \bibfield  {author} {\bibinfo {author} {\bibfnamefont {A.~C.}\ \bibnamefont
  {Hewson}},\ }\href {\doibase 10.1017/CBO9780511470752} {\emph {\bibinfo
  {title} {{The Kondo problem to heavy fermions}}}}\ (\bibinfo  {publisher}
  {Cambridge University Press},\ \bibinfo {address} {Cambridge},\ \bibinfo
  {year} {1997})\BibitemShut {NoStop}%
\bibitem [{\citenamefont {Bauer}\ \emph {et~al.}(2007)\citenamefont {Bauer},
  \citenamefont {Oguri},\ and\ \citenamefont {Hewson}}]{Bauer2007Nov}%
  \BibitemOpen
  \bibfield  {author} {\bibinfo {author} {\bibfnamefont {J.}~\bibnamefont
  {Bauer}}, \bibinfo {author} {\bibfnamefont {A.}~\bibnamefont {Oguri}}, \ and\
  \bibinfo {author} {\bibfnamefont {A.~C.}\ \bibnamefont {Hewson}},\ }\bibfield
   {title} {\enquote {\bibinfo {title} {{Spectral properties of locally
  correlated electrons in a Bardeen{\textendash}Cooper{\textendash}Schrieffer
  superconductor}},}\ }\href {\doibase 10.1088/0953-8984/19/48/486211}
  {\bibfield  {journal} {\bibinfo  {journal} {J. Phys.: Condens. Matter}\
  }\textbf {\bibinfo {volume} {19}},\ \bibinfo {pages} {486211} (\bibinfo
  {year} {2007})}\BibitemShut {NoStop}%
\bibitem [{\citenamefont {Sherman}\ \emph {et~al.}(2016)\citenamefont
  {Sherman}, \citenamefont {Yodh}, \citenamefont {Albrecht}, \citenamefont
  {Nyg{\aa}rd}, \citenamefont {Krogstrup},\ and\ \citenamefont
  {Marcus}}]{Sherman2016Nov}%
  \BibitemOpen
  \bibfield  {author} {\bibinfo {author} {\bibfnamefont {D.}~\bibnamefont
  {Sherman}}, \bibinfo {author} {\bibfnamefont {J.~S.}\ \bibnamefont {Yodh}},
  \bibinfo {author} {\bibfnamefont {S.~M.}\ \bibnamefont {Albrecht}}, \bibinfo
  {author} {\bibfnamefont {J.}~\bibnamefont {Nyg{\aa}rd}}, \bibinfo {author}
  {\bibfnamefont {P.}~\bibnamefont {Krogstrup}}, \ and\ \bibinfo {author}
  {\bibfnamefont {C.~M.}\ \bibnamefont {Marcus}},\ }\bibfield  {title}
  {\enquote {\bibinfo {title} {{Normal, superconducting and topological regimes
  of hybrid double quantum dots}},}\ }\href {\doibase 10.1038/nnano.2016.227}
  {\bibfield  {journal} {\bibinfo  {journal} {Nat. Nanotechnol.}\ }\textbf
  {\bibinfo {volume} {12}},\ \bibinfo {pages} {212} (\bibinfo {year}
  {2016})}\BibitemShut {NoStop}%
\bibitem [{\citenamefont {W{\ifmmode\acute{o}\else\'{o}\fi}jcik}\ and\
  \citenamefont {Weymann}(2018)}]{KWIW-S2QD}%
  \BibitemOpen
  \bibfield  {author} {\bibinfo {author} {\bibfnamefont {K.~P.}\ \bibnamefont
  {W{\ifmmode\acute{o}\else\'{o}\fi}jcik}}\ and\ \bibinfo {author}
  {\bibfnamefont {I.}~\bibnamefont {Weymann}},\ }\bibfield  {title} {\enquote
  {\bibinfo {title} {{Interplay of the Kondo effect with the induced pairing in
  electronic and caloric properties of T-shaped double quantum dots}},}\ }\href
  {\doibase 10.1103/PhysRevB.97.235449} {\bibfield  {journal} {\bibinfo
  {journal} {Phys. Rev. B}\ }\textbf {\bibinfo {volume} {97}},\ \bibinfo
  {pages} {235449} (\bibinfo {year} {2018})}\BibitemShut {NoStop}%
\bibitem [{\citenamefont {Ternes}(2015)}]{Ternes15}%
  \BibitemOpen
  \bibfield  {author} {\bibinfo {author} {\bibfnamefont {M.}~\bibnamefont
  {Ternes}},\ }\bibfield  {title} {\enquote {\bibinfo {title} {{Spin
  excitations and correlations in scanning tunneling spectroscopy}},}\ }\href
  {\doibase 10.1088/1367-2630/17/6/063016} {\bibfield  {journal} {\bibinfo
  {journal} {New J. Phys.}\ }\textbf {\bibinfo {volume} {17}},\ \bibinfo
  {pages} {063016} (\bibinfo {year} {2015})}\BibitemShut {NoStop}%
\bibitem [{\citenamefont {Kahn}(1993)}]{Kahn_book}%
  \BibitemOpen
  \bibfield  {author} {\bibinfo {author} {\bibfnamefont {O.}~\bibnamefont
  {Kahn}},\ }\href@noop {} {\emph {\bibinfo {title} {{Molecular magnetism}}}}\
  (\bibinfo  {publisher} {VCH Publishers},\ \bibinfo {address} {New York},\
  \bibinfo {year} {1993})\BibitemShut {NoStop}%
\end{thebibliography}
%

\end{document}